\RequirePackage{ifpdf}
\documentclass[letterpaper]{JHEP3}
\usepackage{amsmath}
\usepackage{epsfig}
\usepackage{subfigure}

\DeclareSymbolFont{matha}{OML}{txmi}{m}{it}
\DeclareMathSymbol{\varv}{\mathord}{matha}{118}

\newcommand{\roughly}[1]{\mathrel{\raise.3ex\hbox{$#1$\kern-0.85em
\lower1ex\hbox{$\sim$}}}}

\newcommand{\lsim}{\roughly<}

\def\endignore{}
\def\ignore #1\endignore{} 

\def\la{{\bigl \langle}}
\def\ra{{\bigr \rangle}}
\def\cA{{\cal A}}

\def\cX{{\cal X}}
\def\cF{{\cal F}}

\def\cL{{\cal L}}
\def\cM{{\cal M}}
\def\cO{{\cal O}}
\def\cN{{\cal N}}

\def\cY{{\cal Y}}
\def\cR{{\cal R}}

\def\cV{{\cal V}}

\def\cZ{{\cal Z}}

\def\BLF{{\scriptscriptstyle BLF}}

\newbox\charbox
\newbox\slabox
\def\slsh#1{{      
        \setbox\charbox=\hbox{$#1$}
        \setbox\slabox=\hbox{$/$}
        \dimen\charbox=\ht\slabox
        \advance\dimen\charbox by -\dp\slabox
        \advance\dimen\charbox by -\ht\charbox
        \advance\dimen\charbox by \dp\charbox
        \divide\dimen\charbox by 2
        \raise-\dimen\charbox\hbox to \wd\charbox{\hss/\hss}
        \llap{$#1$}
}}

\def\exd{{\hbox{d}}}
\def\d{\exd}

\def\bea{\begin{eqnarray}}
\def\eea{\end{eqnarray}}
\def\be{\begin{equation}}
\def\ee{\end{equation}}

\def\ssA{{\scriptscriptstyle A}}
\def\ssB{{\scriptscriptstyle B}}

\def\ssF{{\scriptscriptstyle F}}

\def\ssJ{{\scriptscriptstyle J}}

\def\ssM{{\scriptscriptstyle M}}
\def\ssN{{\scriptscriptstyle N}}
\def\ssP{{\scriptscriptstyle P}}
\def\ssQ{{\scriptscriptstyle Q}}
\def\ssR{{\scriptscriptstyle R}}

\def\ssT{{\scriptscriptstyle T}}

\def\ssV{{\scriptscriptstyle V}}

\def\KK{{\scriptscriptstyle KK}}
\def\SM{{\scriptscriptstyle SM}}
\def\EH{{\scriptscriptstyle EH}}

\def\V{\mathcal{V}}

\def\nn{\nonumber}

\def\d{\mathrm{d}}

\def\({\left(}
\def\){\right)}

\def\loc{{\rm loc}}
\def\pref#1{(\ref{#1})}
\def\eff{{\rm eff}}

\def\d{\mathrm{d}}

\numberwithin{equation}{section}

\title{Self-Tuning at Large (Distances):\\ 4D Description of Runaway Dilaton Capture}

\author{C.P.~Burgess,$^{1,2,3}$ Ross Diener${}^{1,2}$ and M. Williams$^{4}$ \\
$^1$ Physics \& Astronomy, McMaster University,
Hamilton, ON, Canada, L8S 4M1\\
$^2$ Perimeter Institute for Theoretical Physics, Waterloo, ON, Canada N2L 2Y5\\
${}^3$ Division PH\,-TH, CERN, CH-1211, Gen\`eve 23, Suisse\\
$^4$ Instituut voor Theoretische Fysica, KU Leuven,
B-3001 Leuven, Belgium
}

\date{\today}

\abstract {We complete here a three-part study (see also {\tt arXiv:1506.08095} and {\tt arXiv:1508.00856}) of how codimension-two objects back-react gravitationally with their environment, with particular interest in situations where the transverse `bulk' is stabilized by the interplay between gravity and flux-quantization in a dilaton-Maxwell-Einstein system such as commonly appears in higher-dimensional supergravity and is used in the Supersymmetric Large Extra Dimensions (SLED) program. Such systems enjoy a classical flat direction that can be lifted by interactions with the branes, giving a mass to the would-be modulus that is smaller than the KK scale. We construct the effective low-energy 4D description appropriate below the KK scale once the transverse extra dimensions are integrated out, and show that it reproduces the predictions of the full UV theory for how the vacuum energy and modulus mass depend on the properties of the branes and stabilizing fluxes. In particular we show how this 4D theory learns the news of flux quantization through the existence of a space-filling four-form potential that descends from the higher-dimensional Maxwell field. We find a scalar potential consistent with general constraints, like the runaway dictated by Weinberg's theorem. We show how scale-breaking brane interactions can give this potential minima for which the extra-dimensional size, $\ell$, is exponentially large relative to underlying physics scales, $r_\ssB$, with $\ell^2 = r_\ssB^2 e^{- \varphi}$ where $-\varphi \gg 1$ can be arranged with a small hierarchy between fundamental parameters. We identify circumstances where the potential at the minimum can (but need not) be parametrically suppressed relative to the tensions of the branes, provide a preliminary discussion of the robustness of these results to quantum corrections, and discuss the relation between what we find and earlier papers in the SLED program.}

\begin{document}

\section{Introduction}

In this paper we study the very low-energy dynamics of six-dimensional supergravity interacting with two non-supersymmetric, space-filling, codimension-two branes. Our interest is in situations where the back-reaction of the branes breaks a degeneracy of the bulk system and lifts an otherwise flat direction. As in two earlier papers \cite{Companion, Companion2} we focus on systems for which the interactions are weak enough to ensure that the energetics lifting this flat direction are amenable to understanding in the effective 4D theory below the Kaluza-Klein (KK) scale. We compute this low-energy potential explicitly within the classical limit, to identify how it depends on the various parameters describing the underlying UV completion.

To this end we study a specific system of branes interacting through the bosonic fields of chiral, gauged six-dimensional supergravity \cite{NS}. We use this specific theory for two reasons. First, it is known to admit explicit stabilized extra-dimensional solutions --- both without branes \cite{SS} and with them \cite{SLED, GGP, Swirl, MultiBrane, Nemanja6DBH, 6DdS, TimeDep} --- for which gravity competes with flux quantization and brane back-reaction to stabilize the extra dimensions. This makes it a good laboratory for studying in detail how interactions amongst branes and fluxes can compete to shape the extra dimensions while going beyond the restriction to one extra dimension of the well-explored 5D Randall-Sundrum models \cite{RS}. In this motivation one wishes to know whether or not it is possible to achieve dynamically stable extra dimensions that are exponentially large functions of the not-too-large parameters of the fundamental theory.

Second, this system was proposed some time ago \cite{SLED, SLEDrev, LesHouches} (and again recently in more detail \cite{TNCC}) as a concrete laboratory in which to explore whether the interplay between supersymmetry and extra dimensions can help resolve the cosmological constant problem \cite{LesHouches, CCprob}, essentially by having the quantum zero-point fluctuations of the particles we see curve the extra dimensions instead of the four large dimensions explored by cosmologists. In the simplest picture  ordinary particles are localized on the 4D branes and so their quantum fluctuations contribute to the brane tensions, while many of the simplest brane solutions \cite{SLED, GGP} are flat for any value of the tension. In this motivation the issue is to understand how (and whether) the 4D theory captures this special feature of the extra-dimensional picture, and thereby to understand how robustly (and whether) the effective 4D curvature can be suppressed relative to naive expectations.

In the simplest model \cite{SS} flux quantization and gravity drive the system to a supersymmetric ground state with a single flat direction corresponding to a breathing mode with origins in an accidental scaling symmetry generic to the classical supergravity field equations. Brane back-reaction then typically lifts this degeneracy (and generically breaks supersymmetry) leading to a vacuum configuration whose properties involve a competition between inter-brane forces and flux quantization. Because the energy cost of this lifting is often smaller than the Kaluza-Klein (KK) scale it can be understood purely within the low-energy 4D theory, and a puzzle for these systems has been how this low-energy theory `knows' about extra-dimensional flux quantization (as it must if it is to properly reproduce the competition with other effects in the 6D UV completion).

An important part of this story is the ability of the branes to carry localized amounts of the stabilizing external magnetic flux \cite{BLFFluxQ},
\be \label{BLFAction}
 S_{\BLF} \propto \int \cA(\phi) {}^\star F \,,
\ee
where the integral is over the 4D brane world-sheet ${}^\star F$ is the 6D Hodge dual of the 2-form Maxwell field-strength and $\cA$ is a dilaton-dependent coefficient. This is important because the system often responds to perturbations by moving flux onto and off of the branes, since it is energetically inexpensive to change the value of $\phi$. We use the effective theory that captures the low-energy dynamics of this flux in the higher-dimensional theory --- developed in companion papers \cite{Companion, Companion2} --- to work out the effective 4D description provided here, identifying in particular the precise form of the scalar potential that governs the energetics of vacuum determination.

We find the following main results.
\begin{itemize}
\item{\em 4D effective description:} We describe the low-energy 4D effective theory appropriate for physics below the Kaluza-Klein (KK) scale, within which the extra dimensions themselves are too small to be resolved, and show how this reproduces the dynamics of the known cases where the 6D dynamics is explicitly known. We find that the news of flux quantization comes to the low-energy theory by a space-filling 4-form gauge field, $F_{\mu\nu\lambda\rho}$, whose value satisfies general quantization conditions \cite{BP,SP} that are ultimately inherited from the higher-dimensional quantization of Maxwell flux.
\item {\em Dynamics of modulus stabilization:} Most trivially we verify in more detail earlier claims \cite{BLFFluxQ,6DSUSYUVCaps,6DHiggsStab} that (with two transverse dimensions) brane couplings generically do stabilize the size of the transverse dimensions in supersymmetric models, in a manner similar to Goldberger-Wise stabilization \cite{GoldWis} in 5D. They do so because they break the classical scale invariance of the bulk supergravity that prevents the bulk from stabilizing on its own (through {\em eg}\, flux stabilization).
\item {\em Exponentially large dimensions:}
    We show that simple choices for brane-bulk couplings allow the extra dimensions to be stabilized at a size, $\ell$, that is large relative to other microscopic scales, $r_\ssB$, exponentially\footnote{This echoes a similar claim of \cite{BLFFluxQ} but fixes an error made there (see next bullet point) and provides a precise 4D formulation of the mechanism.} in the parameters of the underlying theory --- {\em i.e.} $\ell^2 / r_\ssB^2 = e^{-\varphi} $, so $\ell/ r_\ssB$ can be enormous if $\varphi$ is only moderately large, say $\cO(10)$, and negative.
\item {\em Connection between brane-dilaton couplings and curvature:}
     As has been known for some time \cite{ScaleLzero} there is a strong connection between the strength of brane-dilaton couplings and on-brane curvatures, with vanishing brane-dilaton couplings implying vanishing on-brane curvatures. More recently \cite{Companion2} --- see also \cite{Germans} --- it was found that the absence of dilaton couplings is {\em not} as straightforward as demanding dilaton-independence of the brane tension and BLF coefficient, $\cA(\phi)$, of \pref{BLFAction}, due to the necessity to hold fixed the Maxwell field far from the brane, rather than at the brane position, when deriving the dilaton dependence of the brane. Complete dilaton-independence of the brane action instead turns out to be equivalent to the condition for scale-invariance, despite the presence of the metrics in the Hodge dual of \pref{BLFAction}. Our 4D potential allows us to compute the subdominant size of the curvature as explicit functions of the deviations from scale-invariance, and verify that they reproduce the curvatures found directly within the 6D UV completion.
\item {\em Low-energy on-brane curvature:} We find that the dynamics of modulus stabilization usually also curves the dimensions along the brane world-sheets, and generically does so by an amount commensurate with their tension, $R \sim G_\ssN T$, where $T$ is the brane tension (defined more precisely below) and $G_\ssN$ is Newton's constant for observers living on the brane. For specific parameter regimes the on-brane curvature can be less than this however, being parametrically suppressed relative to the tension.

    In some cases the suppression of $R$ in the near-scale-invariant limit can be regarded as a consequence of the generic runaway present for scale-invariant potentials: weak scale-breaking tends to place minima out at large fields for which the potential is relatively small. In this way it potentially converts Weinberg's no-go theorem \cite{Wbgnogo} from a bug into a feature.
\end{itemize}

Although our personal motivation for studying this system is because of its potential application \cite{SLED, LesHouches} to the cosmological constant problem \cite{LesHouches, CCprob, Wbgnogo}, the ability to stabilize two transverse dimensions with exponentially large size given only moderately large input parameters potentially puts large-extra-dimensional models \cite{ADD} on a similar footing as warped Randall-Sundrum models \cite{RS}.

\subsubsection*{A road map}

We organize our discussion as follows. The following section, \S\ref{sec:HiDSys}, describes the 6D system whose 4D physics is of interest, summarizing the main results explained in more detail in \cite{Companion2}. The purposes of doing so is to show how properties of the bulk physics (such as extra-dimensional size and on-brane curvature) are constrained by the field equations, which controls the extent to which they depend on the properties of any source branes. This provides the tools required for matching to the 4D effective theory, relevant to energies below the KK scale. This matching is itself described in \S\ref{sec:4DEFT}, which determines the 4D effective theory required to reproduce the dynamics of the full higher-dimensional theory.

Next, \S\ref{sec:STMicro} uses this effective description to explore the implications of several choices of parameters within a class that minimize the couplings between the brane and the bulk dilaton. In particular we compute here the classical predictions for the modulus mass and vev (and so also the size of the extra dimensions) as well as the on-brane curvature at the minimum. We find examples that produce exponentially large dimensions and with parametrically suppressed curvature in the on-brane directions. \S\ref{sec:STMicro} concludes with a brief discussion about the robustness of the various examples, and surveys some ways that quantum corrections might be expected to complicate the picture.
Our conclusions are summarized in a final discussion section, \S\ref{section:discussion}.

\section{The higher-dimensional system}
\label{sec:HiDSys}

We here briefly outline the action and field equations of the UV theory whose low-energy description we wish to capture: the system studied in \cite{Companion2} consisting of a bulk Einstein-Maxwell-Dilaton sector that arises as the bosonic part of six-dimensional supergravity, plus two space-filling 3-branes situated within two transverse extra dimensions.

\subsection{The Bulk}
\label{sec:bulksystem}

The bulk action is a subset of the action for Nishino-Sezgin supergravity \cite{NS} given by
\bea \label{SB}
 S_\ssB &=& - \int \exd^{6}x \; \sqrt{-g} \left[ \frac{1}{2\kappa^2} \; g^{\ssM\ssN} \Bigl( \cR_{\ssM \ssN} + \partial_\ssM \phi \, \partial_\ssN \phi \Bigr) + \frac{2g_\ssR^2}{\kappa^4} \, e^\phi  + \frac14  e^{-\phi}  A_{\ssM \ssN} A^{\ssM \ssN}  \right] \nn\\
 &=:& - \int \exd^{6}x \; \sqrt{-g} \; \Bigl( L_\EH + L_\phi + L_\ssA \Bigr)
 \,,
\eea
where\footnote{We use Weinberg's curvature conventions \cite{Wbg}, which differ from those of MTW \cite{MTW} only by an overall sign in the definition of the Riemann tensor.} $\kappa$ denotes the 6D gravitational coupling and $\cR_{\ssM\ssN}$ denotes the 6D Ricci tensor while $A_{\ssM \ssN} = \partial_\ssM A_\ssN - \partial_\ssN A_\ssM$ is a gauge field strength for a specific $U(1)_\ssR$ symmetry that does not commute with 6D supersymmetry (with gauge coupling $g_\ssR$). The second line sets up notation for the Einstein-Hilbert, scalar and gauge parts of the action in terms of the items in the line above.

Notice $S_\ssB$ scales homogeneously, $S_\ssB \to s^{2} S_\ssB$ under the rigid rescalings $g_{\ssM \ssN} \to s \; g_{\ssM \ssN}$ and $e^\phi \to s^{-1} e^\phi$, making this a symmetry of the classical equations of motion. Besides ensuring classical scale invariance this also shows that it is the quantity $e^{2\phi}$ that plays the role of $\hbar$ in counting loops within the bulk part of the theory.

The bulk system enjoys a second useful scaling property: physical properties depend only on $g_\ssR$ through a field-dependent combination $\hat g_\ssR (\phi) = g_\ssR \, e^{\phi /2}$. The value $\phi = 0$ can always be chosen as the present-day vacuum provided the values of $g_\ssR$ is chosen appropriately.

For many purposes it is useful to work with a 4-form field strength, $F_{\ssM\ssN\ssP\ssQ}$ that is dual to $A_{\ssM\ssN}$, in terms of which the bulk action can be written
\bea \label{SBdual}
 S_\ssB &=& - \int \exd^{6}x \; \sqrt{-g} \left[ \frac{1}{2\kappa^2} \; g^{\ssM\ssN} \Bigl( \cR_{\ssM \ssN} + \partial_\ssM \phi \, \partial_\ssN \phi \Bigr) + \frac{2g_\ssR^2}{\kappa^4} \, e^\phi \right. \nn\\
  && \qquad\qquad\qquad\qquad\qquad\qquad \left. + \frac1{2\cdot 4!} \, e^{\phi} \,  F_{\ssM\ssN\ssP\ssQ} F^{\ssM\ssN\ssP\ssQ} + L_{st} \right] \nn\\
  &=:& - \int \exd^{6}x \; \sqrt{-g} \; \Bigl( L_\EH + L_\phi + L_\ssF + L_{st} \Bigr) \,,
\eea
where $L_{st}$ is a surface term \cite{Companion2, BP} that emerges when performing the duality transformation from $A_{(2)}$ to $F_{(4)} = \exd V_{(3)}$,
\be
  \cL_{st} := \frac{1}{3!} \, \partial_\ssM \Bigl( \sqrt{-g} \;  \epsilon^{\ssM\ssN\ssP\ssQ\ssR\,\ssT} V_{\ssN\ssP\ssQ} A_{\ssR\,\ssT} \Bigr) \,.
\ee

\subsection{The Branes}
\label{sec:branesystem}

We take the brane action to include the first two terms in a derivative expansion\footnote{The quantity $T$ here is denoted $\check T$ in \cite{Companion2}.}
\bea \label{eq:Sbrane}
 S_\eff &=& -\sum_{\varv} \int_{x = z_\varv(\sigma)} \d^4 \sigma \sqrt{-\gamma} \left[ T_\varv(\phi) - \frac{1}{ 4!} \zeta_\varv (\phi) \, \varepsilon^{\mu\nu\lambda\rho} F_{\mu\nu\lambda\rho} \right] \nn\\
 &=:& \sum_{\varv} \int_{z_\varv} \exd^4 \sigma \; \Bigl( \cL_{\varv}^T + \cL_\varv^{\zeta} \Bigr) = \sum_\varv S_\varv \,,
\eea
where the tension term, $\cL_\varv^T$, is built from the induced metric $\gamma_{\mu \nu}(\sigma) = g_{\ssM\ssN} \partial_\mu z_\varv^\ssM \partial_\nu z_\varv^\ssN$ at the position of the brane (with $z_\varv^\ssM(\sigma)$ denoting the brane position fields). Despite its appearances, the localized-flux term, $\cL_\varv^\zeta$, does not depend on this metric because the explicit dependence cancels with that hidden within the totally antisymmetric 4-tensor, $\varepsilon^{\mu \nu\lambda \rho}$, associated with the metric. Since it turns out the branes repel one another their position modes are massive enough to be integrated out in the 4D effective theory, and so we simply assume static branes and choose coordinates so that they are located at opposite ends of the transverse extra dimensions.

It is also possible to frame the branes using a more UV-complete theory for which they arise as classical vortex-like solutions (as is done explicitly in \cite{Companion, Companion2}), though we do not need the details of this explicit extension in what follows.

\subsection{Bulk geometry and field equations}
\label{subsec:ansatz}

Our interest is in geometries that are maximally symmetric in 4D (spanned by coordinates $x^\mu$) and axially symmetric in the transverse 2D (spanned by $y^m$) about the positions of two source branes situated at opposite ends of a compact transverse space. We therefore specialize to fields that depend only on the proper distance, $\rho$, from the points of axial symmetry, and assume the only nonzero components of the gauge field strength, $A_{mn}$, lie in the transverse two directions, and so its dual, $F_{\mu\nu\lambda\rho}$, lies entirely in the space-filling 4D. The metric has the general warped-product form
\be \label{productmetric}
 \exd s^2 = g_{\ssM \ssN} \, \exd x^\ssM \exd x^\ssN = g_{mn} \, \exd y^m \exd y^n + W^2(y) \, \check g_{\mu\nu}(x) \, \exd x^\mu \exd x^\nu \,,
\ee
where $\check g_{\mu\nu}(x)$ is the maximally symmetric metric on $d$-dimensional de Sitter, Minkowski or anti-de Sitter space. The corresponding 6D Ricci tensor has components
\be
 \cR_{\mu\nu} = \check R_{\mu\nu} + g^{mn} \Bigl[ 3\, \partial_m W \partial_n W + W \nabla_m \nabla_n W \Bigr] \, \check g_{\mu\nu} \,,
\ee
and
\be \label{cR2vsR2}
 \cR_{mn} = R_{mn} + \frac{4}{W} \; \nabla_m  \nabla_n W \,,
\ee
where $\nabla$ is the 2D covariant derivative built from $g_{mn}$ and $\check R_{\mu\nu}$ and $R_{mn}$ are the Ricci tensors for the metrics $\check g_{\mu\nu}$ and $g_{mn}$. For the axially symmetric 2D metrics of interest we make the coordinate choice
\be \label{xdmetric}
  g_{mn} \, \exd y^m \exd y^n  =  \exd \rho^2 + B^2(\rho) \, \exd \theta^2 \,.
\ee

With the assumed symmetries the nontrivial components of the matter stress-energy are
\be \label{Tmunusymform}
 T_{\mu\nu}  = - g_{\mu\nu} \; \varrho \,, \qquad
 {T^\rho}_\rho = \cZ - \cX  \qquad \hbox{and} \qquad
 {T^\theta}_\theta = -( \cZ + \cX )  \,,
\ee
where all three quantities, $\varrho = \frac14 \, g^{\mu\nu} T_{\mu\nu}$, $\cX = - \frac12 \, g^{mn} T_{mn}$ and $\cZ$ can be split into bulk and localized brane contributions: $\varrho = \check \varrho_\ssB + \varrho_\loc$, $\cX = \check \cX_\ssB + \cX_\loc$ and so on.\footnote{The localized parts in this split contain all of the terms that carry localized stress-energy, which involves a subtlety when the brane-localized flux parameter, $\zeta$, is nonzero \cite{Companion, Companion2}. The quantities $\check \varrho_\ssB$, $\check \cX_\ssB$ and $\check L_\ssA$ used here are identical to those in \cite{Companion2} and do not contain any localized contributions.} The bulk contributions to these quantities are given by
\be \label{bulkvarrho}
 \check \varrho_{\ssB} = \frac{(\phi')^2}{2\kappa^2} + V_\ssB + \check L_{\ssA} \,,
  \qquad \check \cX_\ssB = V_\ssB + L_\ssF = V_\ssB - \check L_\ssA \qquad \hbox{and} \qquad
 \cZ_\ssB = \frac{(\phi')^2}{2\kappa^2} \,.
\ee

Under the above assumptions the field equations simplify to coupled nonlinear ordinary differential equations. Denoting differentiation with respect to proper distance, $\rho$, by primes, the dilaton field equation reads
\be \label{Bdilatoneom2}
 \frac{1}{BW^4} \, \Bigl( BW^4 \, \phi' \Bigr)' = \kappa^2 \, \left( \cX + \cY \right) \,,
\ee
where the the above equality defines $\cY$. Two things are important about $\cY$: ($i$) $\cY$ contains no terms from the bulk lagrangian and so vanishes identically in the absence of the source branes; and $(ii)$ the brane contribution to $\cY$ vanishes everywhere if and only if the brane lagrangian does not break the scale invariance of the bulk action.

Similarly the three nontrivial components of the trace-reversed bulk Einstein equations reduce to the 4D and 2D trace equations,
\be \label{BavR4-v1}
 g^{\mu\nu} \cR_{\mu\nu} = \frac{\check R}{W^2} + \frac{4}{BW^4} \Bigl( BW' W^{3} \Bigr)' = -2\kappa^2 \cX \,,
\ee
and
\be \label{BavR2}
 g^{mn} \cR_{mn} = R + 4 \left( \frac{W''}{W} + \frac{B'W'}{BW} \right) = - 2 \kappa^2 \left(  \varrho - \frac{\cX}2 \right) \,,
\ee
as well as the $(\rho\rho)$ -- $(\theta\theta)$ equation
\be \label{BnewEinstein}
 \frac{B}{W} \left( \frac{W'}{B} \right)' = - \frac{\kappa^2 \cZ}{2}  \,.
\ee
Notice the special feature of codimension-two sources that eq.~\pref{BavR4-v1} governing the 4D curvature $\check R$ does not depend on the 4D part of the stress-energy, $\varrho$.

\subsection{Brane stress energies}
\label{sec:BSEnergies}

The integrated localized contributions to the stress energy and to $\cY$ can be written as sums over each brane of known functions of the brane tension, $T_\varv$, and localized flux, $\zeta_\varv$. For instance, the energy density is given by
\be \label{eq:rhovsWT}
 \langle \varrho_\loc \rangle = \sum_\varv \varrho_\varv = \sum_\varv W^4_\varv \, T_\varv \,,
\ee
where $W_\varv$ is the metric warp-factor evaluated at the corresponding brane position and we define the notation
\be \label{eq:anglebrack}
 \Bigl\langle \cdots \Bigr\rangle := \frac{1}{\sqrt{-\check g}} \int \exd^2y \, \sqrt{-g} \; \Bigl( \cdots \Bigr) = 2\pi \int \exd\rho \, B W^4 \, \Bigl( \cdots \Bigr) \,.
\ee
It may happen that $W_\varv$ --- or $\phi_\varv$, if $T_\varv = T_\varv(\phi_\varv)$ --- vanishes or diverges at the brane positions, but if so eq.~\pref{eq:rhovsWT} shows this can be absorbed into a renormalization of $T_\varv$ \cite{6DHiggsStab, ClassRenorm}, such as would be expected physically if the value of $T_\varv$ were to be inferred from a measurement of (say) a defect angle, whose size is governed by by the physical energy $\varrho_\varv$. This is addressed in more detail in Appendix~\ref{app:Linearizations}.

Similarly the scale-breaking brane contributions to the dilaton equation are given by
\be
  \langle \cY \rangle = \sum_\varv \cY_\varv \,,
\ee
with
\be \label{Yvsbraneaction}
 \cY_\varv = \frac{W^4_\varv}{2\pi} \left( T_\varv^\prime(\phi) - \frac{1}{4!} \zeta'_\varv(\phi) \epsilon^{\mu \nu \lambda \rho} F_{\mu \nu \lambda \rho} \right) = - \sum_\varv \frac{1}{2\pi \sqrt{-\check g}} \left( \frac{\delta S_\varv}{\delta \phi} \right) \,.
\ee
The brane does not break scale invariance if both the tension and localized flux are independent of $\phi$: $T_\varv' = \zeta_\varv' = 0$. Again, any singularities associated with the vanishing or diverging of fields near the branes can be renormalized into the bulk-brane effective couplings.

The off-brane components of the brane stress-energy are somewhat more subtle to obtain since the dependence of the brane action on the extra-dimensional metric is often only given implicitly. In general, however, stress-energy conservation and the equilibrium balancing of stress-energy within any localized brane ensures these are given by \cite{Companion, Companion2}
\be \label{eq:braneconstraint}
  \langle \cZ_\loc \rangle = \sum_\varv \cZ_\varv \qquad \hbox{and} \qquad
  \langle \cX_\loc \rangle = \sum_\varv \cX_\varv \,,
\ee
with
\be \label{eq:vortexconstraint}
 \kappa^2 \cZ_\varv \simeq \kappa^2 \cX_\varv \simeq - \frac{\kappa^4 \cY_\varv^2}{ 4 \pi} \,,
\ee
where the approximation is valid up to terms that are suppressed by at least two powers of the assumed small ratio between the size of the brane and the size of the bulk.

\subsection{Flux quantization}
\label{sec:fluxquantization}

The symmetry ansatz requires the 4-form field to satisfy
\be
 F_{\mu\nu\lambda\rho} = Q \, \epsilon_{\mu\nu\lambda\rho} \,,
\ee
with $Q$ independent of the 4 space-filling coordinates. The Bianchi identity, $\exd F = 0$, then implies $Q$ also cannot depend on the transverse two coordinates and so is a constant. This constant is the integration constant we would have found if we had explicitly solved the Maxwell field equation for $A_{mn}$.

The value of $Q$ is fixed by flux quantization \cite{Companion2} as follows
\be \label{fluxqn}
 Q = \frac{1}{\widehat \Omega_{-4}} \left[ \frac{2\pi N}{g_\ssR} + \sum_\varv \zeta_\varv(\phi_\varv) \right] \,,
\ee
where $N$ is the integer measuring the total flux of $A_{\ssM\ssN}$ through the transverse two dimensions, and $\zeta_\varv$ is the parameter in \pref{eq:Sbrane} that measures the amount of this flux that is localized onto the position of the brane. Here, $\phi_\varv$ denotes the value of the dilaton at this brane position, and
\be \label{Omegakdef}
 \widehat \Omega_k  := \int \exd^2y \, \sqrt{g_2} \, W^{k}\,  e^\phi  = \int \d^2 y \, \sqrt{\hat g_2} W^k \,,
\ee
represents the integral of $W^k$ over the transverse dimensions using the scale-invariant metric, $\hat g_{mn} := e^\phi \, g_{mn}$, so the particular case $k = 0$ gives the extra-dimensional volume, $\widehat \Omega := \widehat \Omega_0$, as measured by this metric.

\subsection{Boundary conditions}

The near source behaviour of the bulk fields is controlled by the properties of the brane sources, and this manifests in boundary conditions that must be satisfied by the bulk fields as they approach the branes. In practice, these boundary conditions can be derived by integrating the field equations over the localized region containing the brane source, as described in \cite{Companion, Companion2} in more detail. Performing this operation on \pref{Bdilatoneom2}, for example, gives the following boundary conditions for the dilaton at the positions of the branes
\be
 B_\varv W_\varv^4 \phi^\prime_\varv = \frac{ \kappa^2 }{2 \pi} \left( \cX_\varv + \cY_\varv \right) \simeq \frac{ \kappa^2 \cY_\varv }{2 \pi} \,,
\ee
where the approximation uses \pref{eq:vortexconstraint} to identify $\cY_\varv$ as the leading contribution to this boundary condition. Similarly,
\be \label{eq:defect}
1 - W_\varv^4 B^\prime_\varv  =  \frac{ \kappa^2 }{2 \pi} \left( \varrho_\varv  - \cZ_\varv - \frac{1}{2} \cX_\varv \right) \simeq \frac{\kappa^2}{2 \pi} \varrho_\varv = \frac{\kappa^2 W_\varv^4 T_\varv}{2 \pi} \,,
\ee
where the suppression of $\cX_\varv$ and $\cZ_\varv$ implies they are subdominant to the energy density $\varrho_\varv = W_\varv^4 T_\varv.$ Lastly, the boundary condition for the warping in the metric is given by
\be
  B_\varv \left( W_\varv^4 \right)^\prime  = - \frac{\kappa^2 \cX_\varv }{\pi} \,.
\ee
Here and above, a $\varv$ subscript on a buk field (or its derivative) denotes that this quantity is evaluated at the brane position $\rho = \rho_\varv$.

\subsection{Control of approximations}
\label{subsec:scales}

Because we explore classical behaviour it is important to specify its domain of validity. The fundamental parameters of the problem are the gravitational constant, $\kappa$; the gauge coupling, $\hat g_\ssR(\varphi) = g_\ssR e^{\varphi/2}$; and the size of the brane tensions, $T_\varv$, and flux-localization parameters, $\zeta_\varv$.

In the exact, scale invariant solutions of Appendix~\ref{app:SIsolns} the size of the transverse dimensions, $\ell$, can be written in terms of parameters of the lagrangian and the ambient value of dilaton,  $\varphi$, as follows
\be
\ell = ( \kappa  / 2 g_\ssR ) e^{-\varphi/2}  = \kappa / 2 \hat g_\ssR \,.
\ee
In these solutions, the flux integration constant introduced above is given by $Q = 2 g_\ssR /\kappa^2 $ and we use this as a benchmark value when making various estimates.

Weak gravitational response to the energy density of the brane requires $\kappa^2 T_\varv \ll 1$, and this ensures physical observables such as defect angles are small. Similarly, the response to localized flux is controlled by $\kappa^2 Q \zeta^\prime \sim g_\ssR \zeta^\prime $ and so requires $g_\ssR \zeta^\prime \ll 1.$

Since our interest is in the regime where the intrinsic brane width is much smaller than the transverse dimensions we assume throughout $\ell \gg \hat r_\ssV$ where $\ell$ ($ \hat r_\ssV$) is a measure of the extra-dimensional (brane) size. This is accomplished if $\hat r_\ssV / \ell \sim (\hat r_\ssV g_\ssR /  \kappa )e^{\varphi/2}  \ll 1$ which can usually be ensured by requiring
\be
 e^{\varphi} \ll 1 \,,
\ee
although we discuss below an example where the brane size also depends on the value of the dilaton, thus complicating this argument.

Finally, in supergravity semiclassical reasoning also depends on $\varphi$ because it is $e^{2\varphi}$ that counts loops in the bulk theory. Consequently we also require $e^\varphi \ll 1$ in order to work semiclassically.

\subsection{Integral relations}
\label{sec:integrals}

From the point of view of the low-energy theory, it is the field equations integrated over the extra dimensions that carry the most useful information.

Integrating the dilaton field equation, \pref{Bdilatoneom2}, over the entire compact transverse dimension gives
\be \label{dilatontotinteq}
  \Bigl \langle \cX + \cY \Bigr\rangle = 0 \,.
\ee
Since integration over the transverse space can be regarded as projecting the field equations onto the zero mode in these directions, \pref{dilatontotinteq} can be interpreted as the equation that determines the value of the dilaton zero-mode and must agree with what is found by varying the potential of the effective 4D theory obtained in later sections. In the absence of the sources this zero mode is an exact flat direction of the classical equations associated with the scale invariance of the bulk field equations and the localized contribution to \pref{dilatontotinteq} expresses how this flat direction becomes fixed when the sources are not scale-invariant.

Integrating the trace-reversed Einstein equation over the entire transverse space leads to
\be \label{XthetathetaeEinsteininttot}
  \left\langle \varrho - \cZ - \frac{\cX}2 \right\rangle = 0 \,,
\ee
and
\be \label{intRdeq1}
  \check R \, \bigl\langle W^{-2} \bigr\rangle = - 2\kappa^2 \bigl\langle \cX \bigr\rangle  = - 4\kappa^2 \bigl\langle \varrho - \cZ \bigr\rangle  \,,
\ee
where the second equality uses \pref{XthetathetaeEinsteininttot}. This again emphasizes that it is the integrated {\em off-source} stress-energy, $\langle \cX \rangle$, that ultimately controls the size of the on-source curvature \cite{ScaleLzero} for generic $\phi$, and that this receives contributions coming from both bulk and brane-localized contributions to the integral. By contrast, using \pref{dilatontotinteq} to evaluate $\langle \cX \rangle$ --- at the specific value of the would-be zero-mode of $\phi$ that minimizes its potential --- gives a result for the curvature that depends only on brane properties:
\be \label{avR4-vdil}
 \check R \, \bigl\langle W^{-2} \bigr\rangle =  2\kappa^2 \bigl\langle \cY \bigr\rangle \,.
\ee

As we see below, the 6D coupling, $\kappa$, is related to its 4D counterpart, $\kappa_4$, by
\be
 \frac{1}{\kappa_4^2} = \frac{\left\langle W^{-2} \right\rangle}{\kappa^2} \,,
\ee
once evaluated at the minimum of the potential for the would-be zero-mode. So \pref{avR4-vdil} shows that the curvature $\check R$ has a size that is equivalent to what would be obtained by a 4D cosmological constant, $U_\star$, of size
\be \label{Ustarval}
 U_\star = \frac12 \, \langle \cX \rangle = \langle \varrho - \cZ \rangle = - \frac12 \, \langle \cY \rangle \,.
\ee
This explicitly relates the size of the potential at its minimum to the size of scale-breaking on the branes.

\subsection{Orders of magnitude}
\label{sec:gravresponse}

Detailed studies of how the bulk solutions depend on the brane parameters in the UV-complete theories of \cite{Companion2} show several kinds of response are possible.

\medskip\noindent{\em Generic case}

\medskip \noindent In the generic situation
\be
 \kappa^2 \cY_\varv \sim \kappa^2 T^\prime_\varv \sim \kappa^2 T_\varv \,,
\ee
is of order the generic size of a gravitational field coming from the brane energy density. When this is true, it follows from the brane constraints in \pref{eq:vortexconstraint} that
\be
 \kappa^2 \cZ_\varv \simeq \kappa^2 \cX_\varv \sim \kappa^4 T_\varv^2 \,,
\ee
and so are suppressed compared to the naive estimate $\kappa^2 T_\varv$ \cite{Companion2}. Eq.~\pref{Ustarval} then shows the resulting 4D curvature corresponds to an effective 4D cosmological constant of order
\be
 U_\star \sim \sum_\varv \cY_\varv \sim \sum T_\varv \,,
\ee
and so is generically of order the brane tension.

\medskip\noindent{\em Scale invariant case}

\medskip \noindent In the scale invariant case we have $T_\varv' = \zeta_\varv' = 0$ and so the quantities $\cY_\varv$ vanish. Eq.~\pref{eq:vortexconstraint} then implies the off-brane components of the brane stress-energies also satisfy
\be \label{eq:xvanish}
 \cZ_\varv \simeq \cX_\varv  \simeq 0\,.
\ee
Lastly, the vanishing of $\cY_\varv$ ensures the same for the 4D curvature:
\be
 \check R = U_\star = 0 \,.
\ee
As we shall see, in the 4D Einstein frame the scalar potential for the dilaton zero-mode, $\varphi$, turns out to be proportional to $\langle \cX \rangle \propto e^{2\varphi}$, so this vanishing of $\langle \cX \rangle$ and $\check R$ is achieved by having the zero-mode run away to $e^\varphi \to 0$ \cite{Wbgnogo}.

\medskip\noindent{\em Decoupling case $T_\varv' = 0$}

\medskip \noindent An intermediate situation is given by the decoupling choice, for which $T_\varv$ is $\phi$-independent but $\zeta_\varv(\phi) $ is not.  In this case $\cY$ is not exactly zero, but should be suppressed because $\cY$ arises purely from the $\phi$-dependence of a derivatively suppressed term in the brane action
\be
 \kappa^2 \cY_\varv \sim \kappa^2  Q \zeta_\varv^\prime \sim g_\ssR \zeta_\varv^\prime \sim \left[ \frac{\kappa \, \zeta_\varv^\prime(\varphi) }{\ell} \right]   \, e^{-\varphi / 2} \,,
\ee
where the last estimate shows how the derivative suppression can be rewritten as a suppression by the size of the extra dimensions. As a consequence we also have suppressions in the off-brane stress-energy components,
\be
 \kappa^2 \cZ_\varv \sim \kappa^2 \cX_\varv \sim g_\ssR^2 \left( \zeta_\varv^\prime \right)^2 \sim \left[ \frac{\kappa\, \zeta_\varv^\prime(\varphi) }{\ell} \right]^2 e^{-\varphi}  \,,
\ee
and the effective cosmological constant corresponding to $\check R$ satisfies
\be \label{eq:Udecouple}
 \kappa^2 U_\star \sim g_\ssR \zeta^\prime_\varv  \sim \left[ \frac{\kappa \,  \zeta_\varv^\prime(\varphi_\star) }{\ell} \right]  \, e^{-\varphi_\star / 2} \,.
\ee
Our goal in the next sections is to reproduce these estimates using a more carefully computed potential for the low-energy 4D effective theory, and to determine the value of the zero mode that ultimately controls the size of these estimates.

\section{EFT below the KK scale}
\label{sec:4DEFT}

Consider next the viewpoint of a lower-dimensional observer with access only below the KK scale. In particular we address the following puzzle. We know in the full $D$-dimensional theory that flux quantization plays a crucial role in determining the $d$-dimensional curvature that would be seen by any observer below the KK scale \cite{BLFFluxQ}. (We know this because it determines $Q$ through \pref{fluxqn}, and this then governs the size of $\check L_\ssA = - L_\ssF$ appearing in $\varrho$ and $\cX$.) But how is this flux-dependence seen by a lower-dimensional observer who cannot resolve the extra dimensions?

The field content naively available in the generic case to the lower-dimensional observer is fairly limited: a massless graviton $g_{\mu\nu}$; massless gauge bosons, one arising from the higher-dimensional gauge field, $A_\mu$, and another, $B_\mu$, arising from the metric due to the unbroken axial rotational invariance of the extra dimensions; and the dilaton zero-mode, $\varphi$, arising due to classical scale-invariance. Although our tale can be told purely using these fields, our interest in practice is in a bulk coming from higher-dimensional supergravity for which additional light particles also exist.

The low-energy field content available in 6D within Nishino-Sezgin supergravity \cite{NS} also includes the `model-independent' axion, $a$, that is dual to the components $C_{\mu\nu}$ of the bulk Kalb-Ramond field, as well as the harmonic part of the extra-dimensional components of the same field, $C_{mn}$. Because the supersymmetry breaking scale in the bulk is also the KK scale these do not appear with superpartners as supermultiplets in the 4D theory. One of these fields, $C_{mn}$, turns out to Higgs the would-be massless gauge boson, $A_\mu$, which then acquires a mass at the KK scale \cite{Susha}.\footnote{The full story is a bit more complicated, with Green-Schwarz cancellation \cite{GreenSchwarz} of gravitational anomalies in 6D \cite{6Danomaly} implying that the massless 4D field is really a mixture of the two gauge fields, $B_\mu$ and $A_\mu$.}

To understand how flux quantization trickles down to the low-energy EFT it is useful to supplement these fields with the 4-form field, $F_{(4)}$, that is dual to $A_{(2)}$. Although this field has trivial dynamics in the low-energy theory, its constant value knows about flux quantization and so can bring the news about it to the lower-dimensional world.

\subsection{Lower-dimensional action}

With these comments in mind we seek that part of the low-energy 4D EFT describing the dynamics of the 4D metric, $g_{\mu\nu}(x)$, the dilaton zero mode, $\varphi(x)$, and the 4-form field strength, $F_{\mu\nu\lambda\rho}$. Because of the appearance of the low-energy scalar we distinguish several important metric frames: the 6D Einstein-frame (EF) metric, $g_{\mu\nu}$, in terms of which the UV theory is formulated; the scale-invariant frame $\hat g_{\mu\nu} = e^\varphi \, g_{\mu\nu}$ which does not transform under the classical scaling symmetry of the UV theory; and the 4D Einstein-frame metric, $\tilde g_{\mu\nu}$, which must be given by
\be \label{framespart1}
 \tilde g_{\mu\nu} \propto e^{-\varphi} g_{\mu\nu} = e^{-2\varphi} \, \hat g_{\mu\nu} \,,
\ee
since this ensures $\tilde g_{\mu\nu} \to s^2 \, \tilde g_{\mu\nu}$ under the scale transformations, as required for the lower-dimensional Einstein-Hilbert term to scale properly. We do {\em not} similarly canonically normalize the zero mode kinetic term in 4D because we wish to keep its transformation property under the classical scaling symmetry: $e^{-\varphi} \to s \, e^{-\varphi}$.

For subsequent applications it is important to get right the proportionality constant in \pref{framespart1}. In particular, we want it to be unity in the present-day vacuum, $\varphi = \varphi_\star$, which we determine below by minimizing the $\varphi$ scalar potential. Having $\tilde g_{\mu\nu}$ and $g_{\mu\nu}$ differ in normalization amounts to a change of units, and so needlessly complicates the dimensional estimate of the size of terms in the low-energy potential. Consequently we use below the following, more precise, version of \pref{framespart1},
\be \label{framespart2}
 \tilde g_{\mu\nu} = e^{-(\varphi - \varphi_\star)} g_{\mu\nu} \,.
\ee

The most general lagrangian for these fields at the two-derivative level can be written
\bea \label{EF4D}
  \cL_4 &=& - \sqrt{- \tilde g} \; \left[ \frac{1}{2\kappa_4^2} \, \tilde g^{\mu\nu} \Bigl( \tilde R_{\mu\nu} + Z_\varphi(\varphi) \, \partial_\mu \varphi \, \partial_\nu \varphi \Bigr) + V_4(\varphi) \right. \nn\\
  && \qquad\qquad\qquad \left. + \frac{1}{2\cdot 4!} \, Z_\ssF(\varphi) \, F_{\mu\nu\lambda\rho} F^{\widetilde{\mu\nu\lambda\rho}} - \frac{1}{4!} \, \xi(\varphi) \, \tilde\epsilon^{\,\mu\nu\lambda\rho} F_{\mu\nu\lambda\rho} \right] + \cL_{st4} \,,
\eea
where tildes on upper indices indicate that they are raised using the inverse metric $\tilde g^{\mu\nu}$, and $\tilde \epsilon^{\mu\nu\lambda\rho}$ is the appropriate volume tensor built from $\tilde g_{\mu\nu}$ (whose nonzero components are $\pm (- \tilde g)^{-1/2}$). The surface term, $\cL_{st4}$, is given by
\be
  \cL_{st4} := \frac{1}{3!} \, \partial_\mu \Bigl( \sqrt{-\tilde g} \; Z_\ssF \check F^{\widetilde{\mu\nu\lambda\rho}} V_{\nu\lambda\rho} \Bigr) \,,
\ee
and is required to the extent there are boundaries (including asymptotic infinity) whose behaviour we wish to track \cite{Companion2, BP}. This last equation uses the definition
\be
 \check F_{\mu\nu\lambda\rho} := F_{\mu\nu\lambda\rho} - \frac{\xi}{Z_\ssF} \, \tilde\epsilon_{\mu\nu\lambda\rho} \,.
\ee

Notice that the equations of motion for the 3-form gauge potential, $\partial_\mu \left( \sqrt{-\tilde g} \; Z_\ssF \check F^{\mu\nu\lambda\rho} \right) = 0$, imply that evaluating $\cL_{st4}$ at a solution gives
\bea
 \Bigl( \cL_{st4} \Bigr)_{\rm on-shell} &=& \frac{Z_\ssF}{4!} \, \sqrt{-\tilde g} \; \check F^{\widetilde{\mu\nu\lambda\rho}} F_{\mu\nu\lambda\rho} = \frac{Z_\ssF}{4!} \, \sqrt{-\tilde g} \; F^{\widetilde{\mu\nu\lambda\rho}} F_{\mu\nu\lambda\rho} - \frac{\xi}{4!} \, \sqrt{-\tilde g} \; \tilde \epsilon^{\mu\nu\lambda\rho} F_{\mu\nu\lambda\rho} \nn\\
 &=& - \sqrt{- \tilde g} \left(- 2L_\ssF - L_\xi \right)  \,.
\eea
Combining this with the above, evaluating the gauge part of the 4D action using the 4-form equations of motion therefore gives
\be \label{4formonshell4D}
 \Bigl( L_{4{\rm form}} \Bigr)_{\rm on-shell} := L_\ssF + L_\xi + L_{st4} = - L_\ssF \,.
\ee

\subsection{Field equations}

The field equations obtained from the EF 4D action \pref{EF4D} are the field equation for the 3-form gauge potential,
\be \label{4Dgaugeeq}
 \partial_\mu \Bigl[ \sqrt{- \tilde g} \; \Bigl(Z_\ssF \, F^{\widetilde{\mu\nu\lambda\rho}} - \xi \, \tilde \epsilon^{\,\mu\nu\lambda\rho}  \Bigr) \Bigr] = 0 \,.
\ee
Writing $F_{\mu\nu\lambda\rho} = f_4 \, \tilde \epsilon_{\mu\nu\lambda\rho}$ shows that $f_4$ is algebraically fixed in terms of an integration constant, $K_4$ and couplings in the lagrangian,
\be
 f_4 = \frac{K_4 + \xi}{Z_\ssF} \,,
\ee
and because of this $F_{(4)}$ does not describe propagating degrees of freedom. In terms of $f_4$ we have $L_\ssF = - \frac12 \, Z_\ssF \, f_4^2$, so evaluating the action using \pref{4formonshell4D} shows that the influence of the 4-form field is to shift the scalar potential of the remaining scalar-tensor theory to
\be
 U(\varphi) := V_4(\varphi) - L_\ssF(\varphi) = V_4(\varphi) + \frac{Z_\ssF}2 \, f_4^2(\varphi) = V_4(\varphi) + \frac{1}{2 Z_\ssF} \Bigl( K_4 + \xi \Bigr)^2 \,.
\ee

The Einstein equations similarly are
\be
 \widetilde R_{\mu\nu} + Z_\varphi \, \partial_\mu \varphi \, \partial_\nu \varphi = - \kappa_4^2  S_{\mu\nu} \,,
\ee
where $S_{\mu\nu} = T_{\mu\nu} - \frac12 \, \tilde g^{\lambda\rho} T_{\lambda\rho} \, \tilde g_{\mu\nu}$ with stress tensor
\be
 T^{\mu\nu} = \frac{Z_\ssF}{3!} \left[ F^{\widetilde{\mu\lambda\rho\kappa}} \, {F^{\tilde\nu}}_{\lambda\rho\kappa} - \frac18 \, \tilde g^{\mu\nu} \, \widetilde{F^2} \right] - V_4 \, \tilde g^{\mu\nu} \,,
\ee
so
\be
 S^{\mu\nu} = \frac{Z_\ssF}{3!} \left[ F^{\widetilde{\mu\lambda\rho\kappa}} \, {F^{\tilde\nu}}_{\lambda\rho\kappa} - \frac38 \, \tilde g^{\mu\nu} \, \widetilde{F^2} \right] + V_4 \, \tilde g^{\mu\nu} \,.
\ee
The traced Einstein equation therefore is
\be \label{4DReq}
 \widetilde R + Z_\varphi \, \left( \widetilde{\partial \varphi} \right)^2 = \kappa_4^2 \left[ \frac{Z_\ssF}{2\cdot 3!} \, F^2 - 4 V_4 \right] = -4\kappa_4^2 \Bigl( V_4 - L_\ssF \Bigr) \,,
\ee
which again shows the effect of the 4-form field is to shift the potential of the scalar field from $V_4$ to $U = V_4 - L_\ssF$.

Finally, the dilaton equation becomes
\bea \label{4Ddilatoneq}
 Z_\varphi \, \widetilde \Box \varphi &=& \kappa_4^2 \Bigl( V_4' + L_\ssF' + L_\xi' +  L_{st4}' \Bigr) \nn\\
 &=& \kappa_4^2 \Bigl( V_4' - L_\ssF' \Bigr)  \,,
\eea
where primes here denote derivatives with respect to $\varphi$. This is again consistent with the replacement $V_4 \to U = V_4 - L_\ssF$. In this argument it may come as a surprise that $\cL_{st4}$ can contribute at all to the field equations for $\varphi$, given that $\cL_{st4}$ is a surface term which therefore should not contribute to equations of motion at all. It is indeed true that because $\cL_{st4}$ is a surface term it can only contribute to the variation of the action with respect to field variations that are nonzero at the boundaries of spacetime. But when evaluating the $\varphi$ potential we first evaluate the lagrangian (and so in particular $\cL_{st4}$) at the solution to the $V_{\mu\nu\lambda}$ equation of motion, and this solution necessarily contributes to the surface terms whenever its field strength satisfies $F_{\mu\nu\lambda\rho} = f_4 \, \tilde\epsilon_{\mu\nu\lambda\rho}$. It is for this reason that $\cL_{st4}$ contributes to the variation of the action with respect to $\varphi$ if $f_4$ depends on $\varphi$ and the variation is made {\em after} $V_{(3)}$ is eliminated as a function of $\varphi$. This is why its presence resolves \cite{BP} paradoxes that would otherwise arise \cite{Duff} when handling 4-form fields.

\subsection{Matching}

Next we try to identify the unknown functions of $\varphi$ in the 4D theory in a way that captures all of the properties of the 6D theory. Since the main focus is on the 4D theory, we adopt in this section (and in the next section) the notation where $g_{\mu\nu}(x)$ (rather than $\check g_{\mu\nu}$) denotes just the $x^\mu$-dependent 4D part of the 6D metric, $g_{\ssM\ssN}(x,y)$, {\em without} the warp factors, $W^2(y)$, in 6D Einstein frame. So (for instance) $\sqrt{-g_6} = \sqrt{-g_4} \; \sqrt{g_2} \; W^4 = \sqrt{-g_4} \; B W^4$.

\subsubsection*{Form field}

We first match the 4-form field, since this is what passes the flux-quantization conditions down to the low-energy theory. The 6D dual Maxwell field equation, integrated over the extra dimensions, for the geometries of interest is
\be
 \partial_\mu \left\{ \sqrt{-g_4} \left[\int \exd^2 y  \left( \frac{B}{W^4} \right)  e^{\phi} F^{\mu\nu\lambda\kappa} - \sum_\varv \zeta_\varv \; \epsilon^{\mu\nu\lambda\kappa} \right] \right\} = 0 \,,
\ee
where warp factors are written explicitly so that 4D indices are raised (and $\epsilon^{\mu\nu\lambda\kappa}$ is built) with the 4D $g^{\mu\nu}$ rather than the 6D version. This is to be compared with its 4D counterpart, derived above in 4D EF,
\be
 \partial_\mu \Bigl[ \sqrt{- \tilde g_4} \; \Bigl(Z_\ssF \, F^{\widetilde{\mu\nu\lambda\rho}} - \xi \, \tilde \epsilon^{\,\mu\nu\lambda\rho}  \Bigr) \Bigr] = \partial_\mu \Bigl[ \sqrt{- g_4} \; \Bigl(Z_\ssF \, e^{2(\varphi- \varphi_\star)} F^{\mu\nu\lambda\rho} - \xi \, \epsilon^{\,\mu\nu\lambda\rho}  \Bigr) \Bigr] = 0 \,,
\ee
where the first equality transforms to 6D EF from 4D EF. Equating coefficients gives
\be \label{ZFeom}
 \widehat Z_\ssF := Z_\ssF e^{2(\varphi-\varphi_\star)} = \int \exd^2y \left( \frac{B}{W^4} \right) e^\phi = \widehat\Omega_{-4}  \,,
\ee
and
\be \label{zetaform}
  \xi(\varphi) = \sum_\varv \zeta_\varv(\phi_\varv) \simeq \sum_\varv \zeta_\varv(\varphi)  \,.
\ee
In the first equality the dilaton evaluated at the brane positions, $\phi_\varv = \phi(y_\varv) = \varphi \, u_0(y_\varv)$, is implicitly expressed in terms of the amplitude, $\varphi$, of the would-be bulk zero-mode. The second, approximate, equality assumes the zero mode $u_0(y)$ to be $y$-independent so that $\phi_\varv = \varphi$ is the same at the position of all branes.

The solution to the 4-form field equation in 4D is given by
\be
  \widehat Z_\ssF \, F^{\mu\nu\lambda\rho} - \xi \, \epsilon^{\,\mu\nu\lambda\rho} = K_4 \, \epsilon^{\,\mu\nu\lambda\rho} \,,
\ee
where $K_4$ is an integration constant. Similarly the solution to the 6D equation, integrated over the transverse space, is
\be
   K_6 \, \epsilon^{\,\mu\nu\lambda\kappa} = \int_{\rm tot} \exd^2y \, \left( \frac{B}{W^4} \right) \left( e^{\phi} F^{\mu\nu\lambda\kappa} \right) - \sum_\varv \zeta_\varv \, \epsilon^{\,\mu\nu\lambda\kappa}
   = \widehat Z_\ssF \,  F^{\mu\nu\lambda\kappa} - \xi \, \epsilon^{\mu\nu\lambda\kappa} \,,
\ee
where $K_6$ is also an integration constant and the second equality uses \pref{ZFeom} and \pref{zetaform}. Comparing these solutions shows $K_6 = K_4$.

But in 6D the Bianchi identity \cite{Companion2} also tells us that
\be
 F_{\mu\nu\lambda\rho} = Q \, \epsilon_{\mu\nu\lambda\rho} \,,
\ee
where 6D flux-quantization requires
\be \label{fluxqn2}
 Q = \frac{1}{\widehat \Omega_{-4}} \left[ \frac{2\pi N}{g_\ssA} - \varepsilon \sum_\varv  e^{(r+1)\phi_\varv} \left( \frac{2\pi n_{b}}{e} \right) \right]
 =:  \frac{ \cN + \xi}{\widehat Z_\ssF} \,,
\ee
where $\cN := 2\pi N/g_\ssR$. This determines $K_4 = K_6 = (\cN + \xi ) - \xi  = \cN$ so that
\be
 F_{\mu\nu\lambda\rho} =  \left( \frac{ \cN + \xi }{\widehat Z_\ssF} \right) \, \epsilon_{\mu\nu\lambda\rho} \,,
\ee
and so brings the news about flux quantization to the lower-dimensional world \cite{BP,SP}. With this choice $L_\ssF$ evaluates in 4D to
\be \label{LFeval}
 L_\ssF(\varphi) = \frac1{2\cdot 4!} \, \widehat Z_\ssF \, F_{\mu\nu\lambda\rho} F^{\mu\nu\lambda\rho} = - \frac{1}{2 \widehat Z_\ssF}  \bigl[  \cN + \xi(\varphi)  \bigr]^2  = - \frac{1}{2 \widehat \Omega_{-4}} \bigl[  \cN + \xi (\varphi) \bigr]^2 \,.
\ee

\subsubsection*{Einstein-Hilbert term}

The 4D Einstein-Hilbert terms dimensionally reduce in the usual way to give
\bea
 \cL_4 &=& -  \frac{1}{2\kappa^2} \,  \sqrt{-g_4} \; g^{\mu\nu}  R_{\mu\nu} \int_{\rm tot} \exd^2y \, \sqrt{g_2} \; W^2 \nn\\
 &=& -  \frac{1}{2\kappa^2} \,  \sqrt{-g_4} \; g^{\mu\nu}  R_{\mu\nu} \, e^{- \varphi} \int_{\rm tot} \exd^2y \, \sqrt{\hat g_2} \; W^2 e^{-\phi+\varphi} \nn\\
 &=& -  \frac{1}{2\kappa^2} \, e^{-\varphi_\star} \sqrt{- \tilde g_4} \; \tilde g^{\mu\nu} \widetilde R_{\mu\nu}  \int_{\rm tot} \exd^2y \, \sqrt{\hat g_2} \; W^2 e^{-\phi+\varphi} \,,
\eea
which uses $\sqrt{g_2} = \sqrt{\hat g_2}\; e^{-\phi}$ to express things in terms of the scale-invariant 2D measure and we absorb the net zero-mode factor, $e^{-\varphi}$ into the metric when transforming to the 4D EF metric: $\tilde g_{\mu\nu} = e^{-\varphi} g_{\mu\nu}$ (with $\partial \varphi$ terms not written, but handled below). Comparing this with the 4D action gives the following $\varphi$-independent expression for the 4D gravitational coupling,
\be \label{dimredkappa}
 \frac{1}{\kappa_4^2} = \frac{1}{\kappa^2}\, e^{-\varphi_\star} \int_{\rm tot} \exd^2y \; \sqrt{\hat g_2} \; W^{2} e^{-\phi+\varphi} = \frac{2\pi}{\kappa^2} \, e^{\varphi-\varphi_\star} \int_{\rm tot} \exd\rho \; B W^{2}  = \frac{1}{\kappa^2} \, e^{\varphi-\varphi_\star} \left\langle W^{-2} \right\rangle_{\rm tot}  \,.
\ee

Earlier sections remarked on the freedom to shift $\phi \to \phi - \varphi_\star$ in the bulk provided one also rescales coupling constants such as $g_\ssR \to g_{\ssR\star} = g_\ssR \, e^{\varphi_\star/2}$. Eq.~\pref{dimredkappa} reflects this freedom in the following way. If $\phi = 0$ is chosen so that $g_\ssR^2 \lsim \kappa$, then $r_\ssB \sim \kappa/g_\ssR$ is not particularly large so having a large transverse space requires $e^{\varphi_\star} \ll 1$ so that $\ell = r_\ssB \, e^{-\varphi_\star/2} \gg r_\ssB$. In this case \pref{dimredkappa} shows that it is the explicit factor of $e^{-\varphi_\star}$ that makes the 4D Planck mass large compared with the 6D Planck mass. On the other hand if $\phi$ is shifted so that $\varphi_\star \simeq 0$ then we have $g_{\ssR\star}^2 \ll \kappa$ and so $\ell^2 \sim r_{\ssB\star}^2 \gg \kappa$. In this case \pref{dimredkappa} gives a large 4D Planck mass because of the large integration volume, which is of order $\ell^2$ rather than order $\kappa$.

\subsubsection*{Scalar-tensor properties}

To determine the scalar potential and kinetic terms we evaluate the 6D actions at the solution of the 2D metric and 4-form equations of motion, but do {\em not} use the 4D metric or scalar field equations so that these can be kept free. The starting point in 6D is the 2D integral of the 6D EF lagrangian density, which has the form
\be
  \int_{\rm tot} \exd^2y \, \cL_6 = - \int_{\rm tot} \exd^2y \, \sqrt{-g_6} \left[ \frac{1}{2\kappa^2} \, \bigl( \cR_{(4)} + \cR_{(2)} \bigr) + L_{\phi} + L_{\ssF} + L_{st} \right] + \sum_\varv \cL_\varv  \,.
\ee
We first evaluate the 4-form field at the solution to its field equations, using a result proven in \cite{Companion2},
\be \label{6Dggeonshell2}
 \left[ - \int \exd^2 y \sqrt{-g} \Bigl( L_\ssF + L_{st} \Bigr) + \sum_\varv \cL^\zeta_\varv \right]_{\ssF\,{\rm eq}} = + \int \exd^2 y \sqrt{-g} \; L_\ssF = - \int \exd^2 y \sqrt{-g} \; \check L_\ssA \,,
\ee
to get
\bea
  \int \exd^2y \, \Bigl( \cL_6 \Bigr)_{\ssF\,{\rm eq}} &=& - \int_{\rm tot} \exd^2y \, \sqrt{-g_6} \left[ \frac{1}{2\kappa^2} \bigl( \cR_{(4)} + \cR_{(2)} \bigr) + L_{\phi} + \check L_\ssA \right] + \sum_\varv \cL_\varv^T  \nn\\
  &=& \int \exd^2y \, \sqrt{-g_6} \left\{ \frac{1}{2\kappa^2}  \left[ g^{\mu\nu} \Bigl( \cR_{\mu\nu} + \partial_\mu\phi \, \partial_\nu\phi \Bigr) + \cR_{(2)} \right] + \varrho \right\} \,,
\eea
where we also split the $\phi$ kinetic term into its 4D and 2D parts, and use \pref{bulkvarrho} and \pref{eq:rhovsWT} to trade the remaining terms for $\varrho = \check \varrho_\ssB + \varrho_\loc$. We then eliminate $\cR_{(2)}$ using the field equation \pref{BavR2} to find
\be
 \frac{1}{\sqrt{-g_4}} \int \exd^2y \, \Bigl( \cL_6 \Bigr)_{g_2, \ssF\,{\rm eq}} = - \int \exd^2y \, BW^4 \left[ \frac{1}{2\kappa^2}  g^{\mu\nu} \Bigl( \cR_{\mu\nu} + \partial_\mu\phi \, \partial_\nu\phi \Bigr) + \frac{\cX}2 \right] \,,
\ee
where $\cX = \check \cX_\ssB + \cX_\loc$ with $\check \cX_\ssB = V_\ssB + L_\ssF = V_\ssB - \check L_\ssA$ and $\langle \cX_\loc \rangle = \sum_\varv \cX_\varv$ given as before. In these expressions the combination $\langle L_\ssF \rangle$ is to be regarded as the function of $\varphi$ and flux quanta given by \pref{LFeval}.

These are to be compared with the 4D action evaluated using only the 4-form field equations,
\be \label{EF4D66}
  \Bigl( \cL_4 \Bigr)_{\ssF\,{\rm eq}} = - \sqrt{- \tilde g} \; \left[ \frac{1}{2\kappa_4^2} \, \tilde g^{\mu\nu} \Bigl( \tilde R_{\mu\nu} + Z_\varphi \, \partial_\mu \varphi \, \partial_\nu \varphi \Bigr) + V_4 - L_\ssF \right]  \,,
\ee
in which we are also to regard $L_\ssF$ as the 4D $\varphi$-dependent combination
\be
 L_\ssF(\varphi) = \frac1{2\cdot 4!} \, Z_\ssF \, e^{2\varphi} F_{\mu\nu\lambda\rho} F^{\mu\nu\lambda\rho} = - \frac{1}{2Z_\ssF}  \bigl(  \cN + \xi  \bigr)^2 e^{-2\varphi} = - \frac{1}{2 \widehat \Omega_{-4}} \bigl[  \cN + \xi (\varphi) \bigr]^2 \,.
\ee

The $\varphi$ kinetic term comes partly from the dimensional reduction of the kinetic term for $\phi$ and partly from the kinetic term for the radion, $\ell$, in the 6D Einstein-Hilbert action, and gives \cite{Susha}
\be
  -  \frac{1}{4\kappa_4^2} \,  \sqrt{- \tilde g_4} \; \tilde g^{\mu\nu} \frac{ \partial_\mu S \, \partial_\nu S }{S^2} = - \frac{1}{\kappa_4^2} \,  \sqrt{- \tilde g_4} \; \tilde g^{\mu\nu} \partial_\mu \varphi \, \partial_\nu \varphi \,,
\ee
where $S = e^{-2\varphi} \propto \ell^2 e^{-\phi}$. The total kinetic contribution then is
\be
 \cL_4 = -  \frac{1}{2\kappa^2} \, e^{-\varphi_\star} \sqrt{- \tilde g_4} \; \tilde g^{\mu\nu} \left( \widetilde R_{\mu\nu} + 2 \, \partial_\mu \varphi \, \partial_\nu \varphi \right)  \int_{\rm tot} \exd^2y \, \sqrt{\hat g_2} \; W^2 e^{-\phi+\varphi} \,,
\ee
and so $Z_\varphi = 2$.

The remaining terms determine the scalar potential, which we seek in 4D Einstein frame. On the 6D side we have
\bea \label{Uresult}
 e^{-2(\varphi-\varphi_\star)} U(\varphi) &=& - \frac{e^{-2(\varphi-\varphi_\star)} }{\sqrt{-\tilde g_4}} \, \int \exd^2y \, \Bigl( \cL_6 \Bigr)_{g_2, \ssF\,{\rm eq}} = \frac{1}{\sqrt{- g_4}} \int \exd^2y \, \Bigl( \cL_6 \Bigr)_{g_2, \ssF\,{\rm eq}}  \nn\\
 &=&  \frac12 \, \bigl\langle \cX \bigr\rangle = \frac12 \left\{ \bigl\langle V_\ssB \bigr\rangle + \sum_\varv \cX_\varv - \frac{1}{2\widehat \Omega_{-4}} \bigl[ \cN + \xi (\varphi) \bigr]^2 \right\}\,.
\eea
A check on the normalization comes from the 4D Einstein equation which in Einstein frame states $\tilde R = - 4 \kappa_4^2 U$. This agrees with the above given that it implies $- 4 \kappa_4^2 U = - 2 \kappa_4^2 \langle \cX \rangle e^{2(\varphi- \varphi_\star)}$ while on the other hand
$\kappa^{-2} \langle \cR_{(4)} \rangle = \kappa^{-2} R \langle W^{-2} \rangle = \kappa^{-2} \tilde R \langle W^{-2} \rangle e^{-(\varphi-\varphi_\star)} = \kappa_4^{-2} \tilde R \, e^{-2(\varphi-\varphi_\star)}$ and the 6D field equations state $\langle \cR_{(4)} \rangle = - 2 \kappa^2 \langle \cX \rangle$.

On the 4D side, earlier sections show how the 4-form effectively shifts the effective 4D EF potential from $V_4$ to $U = V_4 - L_\ssF$. Comparing with the 6D result then shows
\bea \label{eq:UofVB}
 U(\varphi) &:=& - \frac{e^{2(\varphi-\varphi_\star)}}{\sqrt{- g}} \, \Bigl( \cL_{4\,{\rm pot}} \Bigr)_{\ssF\,{\rm eq}} = - \frac{1}{\sqrt{-\tilde g}} \, \Bigl( \cL_{4\,{\rm pot}} \Bigr)_{\ssF\,{\rm eq}} \nn\\
 &=&  V_4 - L_\ssF = V_4 + \frac{1}{2 \widehat \Omega_{-4}} \bigl[ \cN + \xi (\varphi) \bigr]^2 \,.
\eea
Although this can be solved for $V_4$ this is less useful than directly working with the total effective potential, $U$.

\subsection{Sources of $\varphi$-dependence within $U$}
\label{phidependence}

Eq.~\pref{Uresult} is one of our main results, since it gives the effective potential whose minimization determines the value of the dilaton zero-mode, $\varphi = \varphi_\star$, and thereby also fixes the size of the extra dimensions, since $\ell^2 = r_\ssB^2 e^{-\varphi_\star}$. The value of the potential at this minimum, $U(\varphi_\star)$ also determines the response of the gravitational field implied when $\varphi$ seeks its minimum in this way.

To make this $\varphi$-dependence more explicit we use $V_\ssB = V_0 \, e^\phi$ (with $V_0 = 2 g_\ssR^2/\kappa^4$) so that
\be \label{Uresult2}
 U(\varphi) = \frac12 \left( V_0 \, \widehat\Omega_4 + \sum_\varv \cX_\varv - \frac{1}{2\widehat \Omega_{-4}} \bigl[ \cN + \xi (\varphi) \bigr]^2 \right) e^{2(\varphi - \varphi_\star)} \,.
\ee
There are four main ways that $\varphi$ enters into this expression.
\begin{itemize}
\item The explicit overall factor of $e^{2\varphi}$.
\item The $\varphi$-dependence of the explicit factors of the flux-localization parameter, $\xi(\varphi) = \sum_\varv \zeta_\varv(\varphi)$.
\item The explicit $\varphi$-dependence of the brane stress-energy parameters, $\sum_\varv \cX_\varv(\varphi)$.
\item Some $\varphi$-dependence potentially enters through the integration volumes $\widehat \Omega_k$. Because $\widehat \Omega_k$ is scale invariant it contains no explicit factors of $\varphi$, but there can be a hidden $\varphi$-dependence because $\widehat \Omega_k$ usually also depends implicitly on $T_\varv$ and $\zeta_\varv$ ({\em eg} through the defect angle, $\alpha_\varv - 1 \propto \kappa^2 T_\varv$) and so inherits any $\varphi$-dependence carried by the brane parameters.
\end{itemize}
We next check several special cases the above potential should reproduce.

\subsubsection*{Scale invariance}

When neither $T_\varv$ nor $\zeta_\varv$ depend on $\phi$ the branes preserve the bulk scale-invariance. In this case all of $\widehat \Omega_k$, $T_\varv$, $\cX_\varv$ and $\xi$ are $\varphi$-independent, so the only dependence on $\varphi$ is the overall factor of $e^{2\varphi}$,
\be
 U(\varphi) = \frac12 \left[ V_0 \widehat \Omega_4 + \sum_\varv \cX_\varv - \frac{1}{2 \widehat \Omega_{-4}} \Bigl( \cN + \xi \Bigr)^2 \right] e^{2(\varphi - \varphi_\star)} \,,
\ee
as would be dictated in general grounds by scale invariance. Although this is always minimized at $U = 0$, unless the square bracket vanishes this is achieved by a runaway to zero coupling, $\varphi \to - \infty$, as required by Weinberg's no-go theorem \cite{Wbgnogo}.

\subsubsection*{Vanishing $U(\varphi)$}
\label{sec:vanish}

Whenever $\cX_\varv$ vanishes (such as happens for the BPS vortices \cite{Companion2} for example) or is negligible, and $V_0 = \frac12 \cV_0^2$ is positive, the quantity $e^{-2\varphi} U(\varphi)$ becomes proportional to a difference of squares and it is simple to enumerate sufficient conditions for it to vanish. In particular
\bea
 U &=&  \frac14 \left\{ \cV_0^2 \; \widehat \Omega_4(\varphi) - \frac{1}{\widehat \Omega_{-4}(\varphi)} \Bigl[ \xi (\varphi) + \cN \Bigr]^2 \right\} e^{2(\varphi - \varphi_\star)} \\
 &=& - \frac{1}{4 \widehat \Omega_{-4}(\varphi)} \left[ \xi (\varphi) + \cN - \cV_0 \, \widetilde \Omega(\varphi) \right]  \left[ \xi (\varphi) + \cN + \cV_0 \, \widetilde \Omega(\varphi) \right] e^{2(\varphi-\varphi_\star)} \,,  \nn
\eea
where $\widetilde \Omega^2 := \widehat \Omega_4 \, \widehat \Omega_{-4}$. This clearly vanishes for all $\varphi$ whenever the functions $\xi (\varphi)$ and $\widetilde \Omega(\varphi)$ are related by
\be \label{difcond1}
 \xi (\varphi) = - \cN \pm \V_0 \, \widetilde \Omega(\varphi) \,,
\ee
for all $\varphi$.  When $\widehat \Omega_k$ and $\widetilde \Omega$ are $\varphi$-independent (which at least requires $T_\varv$ to be independent of $\varphi$) then \pref{difcond1} can only be satisfied for all $\varphi$ if $\xi $ is also $\varphi$-independent, which implies scale invariance.

\subsubsection*{Salam-Sezgin solution}

The Salam-Sezgin solution \cite{SS} described in Appendix~\ref{app:ss} has no sources and so $\xi = T_\varv = \cX_\varv =  0$. It is a supersymmetric solution to 6D supergravity and so $V_0 = 2g_\ssR^2/\kappa^4$ and $\cN = \pm 2\pi/g_\ssR$. The solution is unwarped, $W=1$, so $\widehat \Omega_k = \widehat \Omega_{s} := \pi \kappa^2/g_\ssR^2$ for all $k$. With these choices the scalar potential becomes
\be
 e^{-2(\varphi-\varphi_\star)} \, U = \frac12 \left[ \left( \frac{2g_\ssR^2}{\kappa^4} \right) \widehat \Omega_4 - \frac{\cN^2}{2\widehat \Omega_{-4}} \right] = \frac12 \left[ \left( \frac{2g_\ssR^2}{\kappa^4} \right) \widehat \Omega_{s} - \frac{\cN^2}{2\widehat \Omega_{s}} \right] = \frac12 \left( \frac{2\pi}{\kappa^2} - \frac{2\pi}{\kappa^2} \right) = 0  \,,
\ee
as it should, revealing $\varphi$ as the flat direction.

\subsubsection*{Rugby ball solutions}

We can also investigate the shape of the effective potential when scale invariant branes are added to the system. The rugby-ball solutions presented in Appendix~\ref{app:rugbysoln} are generated by identical, scale-invariant, supersymmetric \cite{AccidentalSUSY} branes, and the potential is expected to vanish in this special case. Explicit solutions are also known when more general scale-invariant branes source the bulk \cite{GGP}, although these solutions generally have bulk fields with nontrivial profiles.

We side-step the technical issues associated with nontrivial warping and dilaton profile and treat both cases simultaneously, by assuming that branes' tension, $T$, and localized flux, $\xi = 2\zeta$, are small enough that we can linearize about the Salam-Sezgin solution (and so also choose flux quantum $\cN = \pm 2\pi/g_\ssR$). This assumption allows us to use the linearized scalar potential \pref{eq:finalpotential} calculated in Appendix~\ref{app:Linearizations}. When specialized to the Salam-Sezgin background around which we are perturbing, it reads
\be
 e^{-2(\varphi - \varphi_\star)} U \simeq \frac12 \sum_\varv \cX_{\varv}  + \frac{2}{\kappa^2} \Bigl( \kappa^2 T + g_\ssR \xi  \Bigr)   \,.
\ee
Above, we have tracked the $\cX_\varv$ contribution to the potential, but the branes are scale invariant, so this quantity is also suppressed as in \pref{eq:xvanish}, and can be neglected. It then follows that the potential vanishes when the branes satisfy
\be
 \kappa^2 T = - g_\ssR \xi \,.
\ee
This is identical to the supersymmetry condition on the branes \cite{AccidentalSUSY}, as expected. Incidentally, when the branes are UV completed as supersymmetric vortices \cite{Companion2} it is also true that the vortex BPS conditions ensure $\cX_\varv = 0$ identically.

When the branes are not supersymmetric, the right-hand size reduces to $2 T$ at linear order when $\xi  = 0$, in agreement with the non-SUSY theory \cite{Companion}. In this case, the resulting potential has the standard runaway form expected for scale-invariant couplings \cite{Wbgnogo}.

\section{Self-tuning under scrutiny}
\label{sec:STMicro}

Now that the tools for computing the dilaton potential are assembled, we can minimize it to explore the size of $e^{\varphi_\star}$ and $U_{\star} = U(\varphi_\star)$ as functions of the microscopic choices (like $T_\varv$ and $\zeta_\varv$) that describe the branes.

\subsection{Implications of $\phi$-independent tension}

We expect special things to happen if we can ensure a small $\phi$ derivative near the branes, since we know the curvature vanishes exactly if $\phi'$ vanishes at both branes \cite{Companion2, ScaleLzero}. This asks the brane lagrangian to be chosen to depend as weakly as possible on $\phi$. The simplest choice is to demand complete $\phi$-independence for both $T_\varv$ and $\zeta_\varv$ for all branes, but although it is true that this leads to solutions with $R = 0$ it also implies scale invariance\footnote{This point is less clear when the brane action is formulated using the original Maxwell field, $A_{(2)}$, rather than $F_{(4)}$, and because of this the equivalence between $\phi$-independence and scale-invariance for brane-localized flux terms was misstated in \cite{TNCC,BLFFluxQ}.} and the results of the previous section confirm that flat curvature in this case is found by having $\varphi$ run away to infinity (thereby not breaking scale invariance) \cite{Wbgnogo}. Consequently in this section we instead choose $\phi$-independence just for the leading term, $T_\varv$, in the hopes that the resulting curvatures can be suppressed.

In this case only two sources of $\varphi$-dependence remain in $U$: the overall factor of $e^{2\varphi}$ and any dependence arising within $\xi (\phi) = \sum_\varv \zeta_\varv(\phi)$. (The latter of these includes both the explicit $\xi$-dependence and any implicit dependence of $\widehat \Omega_k$ on $\xi$.) Because the branes break scale invariance we expect the flat direction for $\varphi$ to be lifted and the dynamics to choose an energetically preferred value, $\varphi_\star$. Furthermore, since the lifting comes from $\xi$, which arises only from the derivatively once-suppressed localized-flux term, we expect $\cY_\varv$ and direct brane contributions to the potential like $\cX_\varv$ to be KK-suppressed --- as argued in more detail in \cite{Companion2}.

This leaves the bulk contribution to $U$, but because of \pref{Ustarval} this is also expected to be suppressed once $\varphi$ adjusts to approach the value $\varphi_\star$. What we do here that \cite{Companion2} did {\em not} do was compute the shape of $U$ explicitly and minimize it to determine $\varphi_\star$ and $U_\star = U(\varphi_\star)$, thereby showing in detail how direct brane contributions to $U$ compete with the interference the branes cause in the cancelations among the bulk terms in $U$.

Because $\ell \propto e^{-\varphi_\star/2}$ in the vacuum this calculation of $\varphi_\star$ also computes the size of the extra dimensions, and we seek solutions with a large hierarchy between the brane size and the size of the transverse dimensions: $\ell \gg \hat r_\ssV$. It is only for such solutions that the above arguments would suggest any suppression in $U_\star$.

\subsubsection*{Consequences of $\partial T/\partial \phi = 0$}

For these reasons our main interest is in situations where $T$ is $\varphi$-independent but $\zeta_\varv = \zeta_\varv(\varphi)$. We next argue that this ensures the contribution of $\cX_\varv$ to $U$ becomes negligible.

Ultimately, it is the derivative suppression of $\zeta$ within the brane action in \pref{Yvsbraneaction} that suppresses $\cX_\varv$ in the potential. For instance, neglecting any $\varphi$-dependence in $\widehat \Omega$ gives the first estimate \cite{Companion2}
\be \label{Yest}
 \cY = \sum_\varv \cY_\varv \sim Q \sum_\varv \zeta_\varv' = Q \, \xi' \simeq \frac{1}{\widehat \Omega_{-4}} \left( \cN + \xi \right) \xi'\,,
\ee
which uses flux quantization to eliminate the bulk flux $Q$. This expression for $\cY$ determines $\sum_\varv \cX_\varv$ through the brane constraint, \pref{eq:vortexconstraint}, which implies
\be
 \sum_\varv \cX_\varv \simeq \frac{\kappa^2 \cY^2}{4\pi} \sim \frac{1}{4\pi} \Bigl( \kappa Q \, \xi' \Bigr)^2 \simeq \frac{1}{4\pi} \left( \frac{\kappa}{\widehat \Omega_{-4}} \right)^2 \Bigl[ \left( \cN + \xi \right) \xi' \Bigr]^2 \,.
\ee
Inserting this information into $U$ then shows that the contribution of $\cX_\varv$ may be dropped relative to the $(\cN + \xi)^2/\widehat \Omega_{-4}$ term whenever $(\kappa \xi')^2 \ll 2\pi \widehat \Omega_{-4}$, as is true when the extra dimensions are much larger than the microscopic sizes determining $\kappa$ and $\xi'$.

As before, for supergravity we have $V_0 = 2g_\ssR^2/\kappa^4$ and as argued above the only $\varphi$ dependence enters through $\xi$ and the overall factor of $e^{2\varphi}$ dictated by scaling, making the scalar potential in the 4D theory
\bea \label{Uvsphisimple}
 U(\varphi) &=& \frac12 \left\{ \left( \frac{2g_\ssR^2}{\kappa^4} \right) \widehat \Omega_{4} (\varphi) - \frac{1}{2 \widehat \Omega_{-4}(\varphi)} \Bigl[ \cN + \xi (\varphi) \Bigr]^2 \right\} e^{2(\varphi -\varphi_\star)} \nn\\
 &=& \frac{1}{4 \widehat \Omega_{-4}} \left(  \frac{2g_\ssR \, \tilde \Omega}{\kappa^2} + \cN  + \xi  \right) \left(  \frac{2g_\ssR \, \tilde \Omega}{\kappa^2} - \cN - \xi  \right) e^{2(\varphi - \varphi_\star)} \,,
\eea
where $\tilde \Omega^2 := \widehat \Omega_4 \widehat \Omega_{-4}$ and in the first line we write $\widehat \Omega_k(\varphi)$ to emphasize that the volumes can also depend on $\varphi$ through $\xi$.

\subsubsection*{General features}

Broadly speaking the potential described above has the form
\be \label{broadform}
 U(\varphi) = F(\varphi) \, e^{2(\varphi-\varphi_\star)} \,,
\ee
and so its extrema, $\varphi_\star$, make the derivative
\be \label{UpgivenF}
 U'(\varphi) = \Bigl( 2F + F' \Bigr) \, e^{2(\varphi-\varphi_\star)} \,,
\ee
vanish. Our interest is in minima, so we demand the second derivative
\be
 U''(\varphi) = \Bigl( 4F + 4F' + F''\Bigr) \, e^{2(\varphi-\varphi_\star)} \,,
\ee
be positive.

There are two classes of solution:
\begin{enumerate}
\item The runaway: $\varphi_\star = \varphi_\infty = - \infty$, with $e^{\varphi_\star} = 0$ and so $U_\star = U_\star'' = 0$; and
\item Any nontrivial solutions to $F'(\varphi_\star) + 2 F(\varphi_\star) = 0$. Evaluated at any of these latter extrema we have\footnote{Notice that the factor of $e^{2\varphi}$ does not suppress $U_\star$ because of the compensating factor of $e^{-2\varphi_\star}$. Although $e^{2\varphi} \propto 1/\ell^4$ ensures the potential is generically suppressed by $1/\ell^4$, the $e^{-2\varphi_\star}$ compensates by converting the prefactor from 6D to 4D Planck density.}
    \be
     U_\star = - \frac12 \, F_\star' \qquad \hbox{and}
     \qquad U_\star'' = 2F_\star' + F_\star'' \,.
    \ee
    Control of approximations requires we check that at any such a minimum $e^{\varphi_\star}$ is small enough to justify our semiclassical analysis.
\end{enumerate}

Our main interest is in the non-runaway minima, and for these notice that using \pref{Uvsphisimple} to infer $F$ and neglecting the $\varphi$-dependence of $\tilde\Omega$ when differentiating the result gives an expression for $U_\star$ that agrees with the estimate of \pref{Yest}. This shows in a more pedestrian way how the low-energy theory knows of the higher-dimensional connection between $U_\star$ and $\langle \cY \rangle$.

Of particular interest is how specific choices for $\zeta_\varv$ (and so also $\xi = \sum_\varv \zeta_\varv$) influence the shape of $F(\varphi)$, and through this the values of $\varphi_\star$ and $U_\star$. We seek to arrange two things: ($i$) that $-\varphi_\star$ be moderately large (to achieve large extra dimensions, given $\ell \propto e^{-\varphi_\star/2}$); and ($ii$) that $U_\star$ be suppressed below the generic brane scale $T_\varv$ (as required to make progress on the cosmological constant problem if ordinary particles are localized on the branes and so contribute their vacuum energies as corrections to the corresponding brane tension).

One way to achieve these ends would be to arrange $F(\varphi) = F_0 \, \cF(\epsilon \varphi)$, where $\epsilon$ is a moderately small dimensionless parameter and $F_0$ is a very small energy density. In this case the linearity of \pref{UpgivenF} ensures the value of $\varphi_\star$ does not depend on $F_0$ at all, and if $\cF(x)$ contains only order-unity parameters we expect to find $|\varphi_\star| \sim \cO(1/\epsilon)$. Having $\varphi_\star \sim -75$ would ensure $e^{-\varphi_\star/2} \sim 10^{16}$; adequate even for models with very large extra dimensions \cite{SLED, ADD}. The question is whether there is enough freedom available in $\xi(\varphi)$ to arrange both of these conditions, and if so whether the choices made can be technically natural.

The next sections explore this question by choosing $\xi = \mu f(\varphi)$ for several simple choices, where $\mu$ is a mass scale that can be adjusted independently from the scale in $T_\varv$. Although we find no obstruction in principle to being able to obtain both large $\varphi_\star$ and small $U_\star$, the simple examples we explore so far each only appear to accomplish one or the other and not both simultaneously.

\subsection{Perturbative solutions}

As argued in \S\ref{sec:vanish}, there are several values of $\varphi$ for which we know $U(\varphi)$ must vanish. One of these is the limit $\varphi \to - \infty$, for which $U \to 0$ because of its exponential prefactor. The second case where we know $U = 0$ is when $\varphi = \varphi_s$ is such that $\xi(\varphi)$ happens by accident to pass through a point where its value agrees with the supersymmetric limit for the given tension. (As shown in the Appendix, at the linearized level this occurs for any $\varphi_s$ satisfying $g_\ssR \xi(\varphi_s) = \mp \kappa^2 T$, if the two branes share equal tensions.) Whenever this occurs $Q$ also takes its supersymmetric value, which ensures $\check R = 0$ (and so $U = 0$).

The significance of such a zero is that it guarantees the existence of at least one maximum or a minimum for $U$ in the range $- \infty < \varphi < \varphi_s$. (A similar conclusion is also possible for any interval between two distinct solutions to $g_\ssR \xi(\varphi_s) = - \kappa^2 T$, should more than one of these exist.) If this extremum is sufficiently close either to $\varphi_\infty$ or to $\varphi_s$ then we can analyze the shape of the potential by perturbing around the situation where $U$ vanishes.

To that end let us write the brane properties as $T_\varv = T_0 + \delta T_\varv$ and $\zeta_\varv = \zeta_0 + \delta \zeta_\varv$, where $T_0$ and $\zeta_0$ define a supersymmetric configuration for which $g_\ssR \xi_0 = g_\ssR \xi(\varphi_s) = 2 g_\ssR \zeta_0 = \mp \kappa^2 T_0$. Then the unperturbed potential vanishes, $U_0 = 0$, and deviations from this can be computed perturbatively in $\delta T_\varv$ and $\delta \zeta_\varv$. There are two naturally occurring small parameters with which to linearize, $\kappa^2 \delta T \ll 1$ and $g_\ssR \delta \xi(\varphi) \ll 1$, whose relative size is a knob we get to dial. Both of these are small to the extent that the bulk is only weakly perturbed by the source branes.

This leads to a potential of the generic form
\be
 U = \Bigl( A + B y + \cdots \Bigr) e^{2(\varphi - \varphi_\star)} \,,
\ee
where $y(\varphi) := g_\ssR \delta \xi/2\pi \ll 1$, and the linearized calculation of the Appendix --- culminating in \pref{eq:finalpotential} --- shows the coefficients $A$ and $B$ are given by
\be
 A \simeq \sum_\varv \delta T_\varv = 2 \, \delta T_{\rm avg} \qquad \hbox{and}
 \qquad B \simeq \frac{4\pi}{\kappa^2}  \,,
\ee
where $\delta T_{\rm avg} = \frac{1}{2} \sum_\varv \delta T_\varv$. Consequently
${A}/{B} \simeq {\kappa^2 \delta T_{\rm avg}}/{2\pi }  \ll 1$.

For this potential
\be
 U' = \Bigl[ 2A + 2 By + B y'  + \cdots  \Bigr] e^{2(\varphi - \varphi_\star)} \,,
\ee
and at non-runaway solutions, $U_\star' = U'(\varphi_\star) = 0$, we have
\be
 U_\star = - \frac12 B y'_\star   + \cdots
 \qquad \hbox{and} \qquad
 U_\star'' =  2 B  y'_\star +  B  y_\star''  + \ldots \,.
\ee

We now describe several types of extrema that such a potential generically possesses. In each case we do not propose an explicit form for $\delta \xi$ for all $\varphi$ (and so also do not compute the potential $U$ for all $\varphi$), but instead investigate its structure near the extrema of $U$ subject to various assumptions about how $\delta \xi$ varies in this region. As a result we do not in these first examples try to compute the value of $\varphi_\star$ from first principles, but only its difference from the position, $\varphi_r$, of a nearby reference point (such as a zero of $U$ or a minimum of $\xi(\varphi)$ {\em etc}). We solve for all quantities in terms of the reference point, $\varphi_r$, and comment on the size of $U_\star$, the KK scale, $\ell$ and the zero-mode mass, $m_\varphi$, at the minimum.

\subsubsection*{Case I: Near a zero of $U$}

Consider first the simplest situation where $\delta \xi$ depends very weakly on $\varphi$ so we may Taylor expand $\xi$ about the point $\varphi = \varphi_s$ where $U$ vanishes
\be
 y(\varphi) \simeq \left( \frac{g_\ssR \mu}{2\pi} \right) \Bigl[ (\varphi - \varphi_s) + \cO\left[ (\varphi- \varphi_s)^2 \right] \Bigr] \,,
\ee
and we assume $|g_\ssR \mu/2\pi| \ll 1$. The potential near $\varphi = \varphi_s$ becomes
\be
 U = \Bigl[ b (\varphi - \varphi_s)  + \cdots \Bigr] e^{2(\varphi - \varphi_\star)} \,,
\ee
where $b \simeq 2 g_\ssR \mu/\kappa^2$.

Extrema are determined by the vanishing of
\be
 U' = \Bigl[ b + 2b ( \varphi - \varphi_s ) + \cdots \Bigr] e^{2(\varphi - \varphi_\star)}  \,,
\ee
and so for finite $\varphi_\star$ this implies
\be \label{eq:phistarv}
 \varphi_\star \simeq \varphi_s - \frac12  \,.
\ee
The condition $g_\ssR \mu/2\pi \ll 1$ ensures that $|y_\star| \ll 1$ at this point, justifying our perturbative analysis of the extremum. The corresponding physical KK scale is
\be \label{ellvsrBI}
 \ell = r_\ssB \, e^{-\varphi_\star/2} = \left( \frac{\kappa}{2g_\ssR} \right) e^{1/4} e^{- \varphi_s / 2}  \,.
\ee

In agreement with \cite{BLFFluxQ, 6DSUSYUVCaps}, the breaking of scale-invariance by the branes allows their back-reaction to stabilize the size of the extra-dimensions, in a 6D version of the Goldberger-Wise \cite{GoldWis} mechanism in 5D. The stabilized size of the extra dimensions is exponentially large compared to microscopic scale $r_\ssB$ to the extent that $\varphi_s$ is large and negative. The full linearization of the 6D system for this example is also given in Appendix~\ref{app:examples}, including a discussion of the warping and dilaton profile generated by the bulk response to the brane perturbations, and of the renormalizations of brane couplings that these require. Later examples also  provide concrete cases for which the value of $\varphi_s$ can be computed in terms of brane properties, and briefly discuss choices that can make $\varphi_s$ large and negative.

At this extremum we have
\be
 U_\star \simeq b  (\varphi_\star - \varphi_s) \simeq - \frac{b}{2} \approx - \frac{g_\ssR \mu}{\kappa^2} \,,
\ee
while
\be
 U_\star'' \simeq  4b + 4b ( \varphi_\star - \varphi_s) \simeq 2 b \approx \frac{4 g_\ssR \mu}{\kappa^2} \,.
\ee
We see we have a local minimum (maximum) between $\varphi = \varphi_s$ and $\varphi \to - \infty$ when $b \propto g_\ssR \mu$ is positive (negative), for which $U_\star$ is negative (positive).\footnote{We are not too concerned here if $U_\star$ turns out negative at the minimum, even for applications to the cosmological constant problem. That is because the goal then is just to have the classical value be {\em smaller} than the inevitable quantum corrections (such as bulk Casimir energies) whose size is hoped to describe the observed (positive) Dark Energy in any ultimately successful model.}

Keeping in mind the normalization of the $\varphi$ kinetic term in the 4D theory we see the classical prediction for its mass at this minimum is
\be
 m^2_\varphi = \frac12\,  \kappa_4^2 \, U_\star'' \simeq 2 \kappa_4^2 \, |U_\star| \simeq \frac{2 g_\ssR \mu}{\left\langle W^{-2} \right\rangle} \,.
\ee
Since generically $\left\langle W^{-2} \right\rangle$ is of order the KK volume we see $m_\varphi$ is suppressed below the KK scale by the small factor $g_\ssR \mu/2\pi$, justifying its calculation in the 4D EFT. This same factor provides the suppression of $U_\star$ relative to the 6D Planck scale, and as a result $m_\varphi^2 \sim |U_\star|/M_p^2$. We return below to a discussion of the robustness of such predictions to quantum corrections.

\subsubsection*{Case II: Near a minimum of $\xi$}

Consider next a situation where $\varphi = \varphi_m$ is a local minimum of $\xi(\varphi)$, and where $\xi_m = \xi(\varphi_m)$ is {\em not} a point where $U$ vanishes. In this case we expand $\xi$ in powers of $\varphi - \varphi_m$ to write $T_\varv = T_0 + \delta T_\varv$ and $\xi = \xi_0 + \mu (\varphi - \varphi_m)^2$. Here $T_0$ is chosen so that $g_\ssR \xi_0 = - \kappa^2 T_0$ (and we choose $N=+1$) so that it is $\delta T_{\rm avg} = \frac12 \sum_\varv \delta T_\varv$ that controls the value of $U$ at $\varphi = \varphi_m$. To justify the perturbative analysis we assume the resulting $\delta T$ satisfies $|\kappa^2 \delta T| \ll 1$ and $|g_\ssR \mu/2\pi| \ll 1$.

With these choices we then have
\be
 y(\varphi) \simeq \left( \frac{g_\ssR \mu}{2\pi} \right) (\varphi - \varphi_m)^2 + \ldots \,,
\ee
and the potential becomes
\be
 U =  \Bigl[ a + b (\varphi - \varphi_m)^2 + \cdots \Bigr] e^{2(\varphi - \varphi_\star)}\,,
\ee
where $a \simeq \sum_\varv \delta T_\varv = 2 \, \delta T_{\rm avg}$ and $b \simeq 2g_\ssR \mu/\kappa^2$. Their dimensionless ratio
\be
 \frac{a}{b} \simeq \frac{\kappa^2 \delta T}{g_\ssR \mu} \,,
\ee
is a free parameter.

The extrema, $\varphi_\star$, are determined by the vanishing of
\be \label{eq:Uprimequad}
 U' \simeq 2\Bigl[ a + b (\varphi - \varphi_m) + b (\varphi-\varphi_m)^2 + \cdots \Bigr]  e^{2(\varphi - \varphi_\star)} \,,
\ee
and so the non-runaway solutions satisfy
\be
 \varphi_{\star\pm} \simeq \varphi_m - \frac12 \left( 1 \pm \sqrt{1 - \frac{4a}{b}} \right) \,.
\ee
Reality of this root requires $4a/b \le 1$ and so $4 \kappa^2 \delta T \le g_\ssR \mu$.

If $|a/b| \ll 1$ the roots take the approximate forms
\be
 \varphi_{\star+} \approx \varphi_m - 1 \,,
 \varphi_{\star-} \approx \varphi_m - \frac{a}{b} \,,
\ee
and if $a/b$ is large and negative they become
\be
 \varphi_{\star\pm} \approx \varphi_m \mp \sqrt{\frac{a}{b}} \,.
\ee
Because $\varphi_{\star-}$ approaches $\varphi_m$ as $a/b \to 0$ perturbation theory also justifies the expansion of $\delta \xi$ in powers of $\varphi - \varphi_m$ for this root. It may nonetheless be justified in any case for the other roots if it happens that $\delta\xi$ remains quadratic out to sufficiently large $\varphi - \varphi_m$, and that $y$ remains small for all of this range.

There are two parameter regimes of interest. The first is $|4a / b| \ll1$ and for this choice $\varphi_\star - \varphi_m$ has the same order of magnitude as $a/b \sim \kappa^2 \delta T /g_\ssR \mu$. As before, the corresponding physical KK scale is $\ell = \left( {\kappa}/{2g_\ssR} \right)  e^{-\varphi_\star/2}$, and because $\varphi_\star - \varphi_m$ is at most order unity, having this be large compared to microscopic scales requires $\varphi_m$ large and negative.

At the extremum $\varphi_\star - \varphi_m \approx - a/b$ we have
\be
 U_\star \simeq a + b  (\varphi_\star-\varphi_m)^2 \simeq - b  (\varphi_\star - \varphi_m) \approx a \simeq 2\, \delta T_{\rm avg} \,,
\ee
while
\be
 U_\star'' =  2[ 2a + b + 4b (\varphi_\star - \varphi_m) + 2 b (\varphi_\star-\varphi_m)^2] \simeq 2b \left[1 +  2( \varphi_\star - \varphi_m)  \right]\approx 2b \simeq \frac{4 g_\ssR \mu}{\kappa^2} \,.
\ee
We see that this is a local minimum when $b \propto g_\ssR \mu$ is positive ({\em ie} whenever $\varphi = 0$ was a minimum for $\delta \xi$). Furthermore the back-reaction with the bulk drags the value of $\varphi_\star$ to be smaller (larger) than the minimum of $\delta \xi$ depending on whether or not $a \simeq 2\delta T_{\rm avg}$ is positive (negative). The value of the potential at this point is $U_\star \approx 2 \delta T_{\rm avg}$ and so is unsuppressed relative to (and shares the same sign as) $\delta T_{\rm avg}$. At this minimum the classical prediction for the would-be zero-mode mass is driven by its potential on the brane,
\be
 m^2_\varphi = \frac12\,  \kappa_4^2 \, U_\star'' \sim \frac{g_\ssR \mu}{\left\langle W^{-2} \right\rangle} \,,
\ee
and so is below the KK scale because $g_\ssR \mu \ll 1$.

Another interesting parameter range enumerated above takes $a / b$ large and negative. In this case
\be
 \varphi_\star = \varphi_m  - \sqrt{\left| \frac{a}{b} \right|} \,,
\ee
provided the quadratic form for $\delta \xi$ applies for fields this large. Notice that $y_\star \sim g_\ssR \mu (\varphi_\star - \varphi_m)^2 \sim a \sim \kappa^2 \delta T_{\rm avg}$ remains small.

Of special interest in this case is where $\sqrt{|a/b|}$ dominates $\varphi_m$, since this could explain why $\varphi_\star$ is also large and negative (and so why $\ell \propto e^{-\varphi_\star / 2}$ could be potentially enormous without needing to explain the size of $\varphi_m$).

At this extremum the size of the potential is
\be
 U_\star \simeq a + b  (\varphi_\star-\varphi_m)^2 \simeq - b  (\varphi_\star - \varphi_m) \approx b \sqrt{\left| \frac{a}{b} \right|} = \sqrt{|ab|} \; (\hbox{sign $b$}) \,,
\ee
which is suppressed relative to the tension scale, $a = 2 \delta T_{\rm avg}$, by the assumed small quantity $\sqrt{|b/a|} \simeq |g_\ssR \mu/\kappa^2 \delta T_{\rm avg}|^{1/2} \ll 1$.  Similarly,
\bea
 U_\star'' &=&  2[ 2a + b + 4b (\varphi_\star - \varphi_m) + 2 b (\varphi_\star-\varphi_m)^2] \nn\\
 &\simeq& 2b \left[1 +  2( \varphi_\star - \varphi_m)  \right] \approx - 4b \sqrt{\left| \frac{a}{b} \right|} = -4 \sqrt{|ab|} \; (\hbox{sign $b$})
 \simeq -4 U_\star \,.
\eea
and the concavity of the potential is once again controlled by $b$, with $b <0$ (and so $a > 0$) giving a minimum at large negative values of $\varphi_\star$. The classical mass of the would-be zero mode at this minimum is
\be
 m_\varphi^2 = \frac{1}{2} \kappa_4^2 U_\star'' \simeq - 2 \kappa_4^2 U_\star \sim \frac{\sqrt{\kappa^2 \delta T |g_\ssR \mu |}}{\la W^{-2} \ra}  \,,
\ee
and this lies below the KK scale beause $|g_\ssR \mu| \ll \kappa^2 \delta T \ll 1$ by assumption. Although this gives large dimensions or small $U_\star$, it does not provide a phenomenologically viable value for both simultaneously, inasmuch as a large-volume value like $\varphi_\star \sim - 75$ only provides a moderate suppression of $U_\star$ relative to tension scales.

The extension of this example to a perturbation in the full 6D theory is also given in Appendix~\ref{app:examples}, including a discussion of brane renormalization.

\subsubsection*{Case III: Near a singular point of $\xi$}

The previous examples assume $\xi$ varies smoothly with $\varphi$, so we next consider a singularity in $\xi$ at $\varphi = \varphi_c$. Singularities can arise in low-energy actions at places in field space where the low-energy approximation fails, such as places where integrated-out species of particles become massless.

For purposes of illustration we consider a branch point of the form $\xi = \xi_0 + \delta \xi = \mu (\varphi - \varphi_c)^\eta$ with $\eta$ an arbitrary exponent. The case $\eta$ near zero is particularly interesting because this profits by being near the scale-invariant case $\eta = 0$. As above, we write $T_\varv = T_0 + \delta T_\varv$ and dial $T_0$ so that it is related to $\xi_0$ by $g_\ssR \xi_0 = - \kappa^2 T_0$.

The potential becomes
\be
 U =  \Bigl[ a + b (\varphi - \varphi_c)^\eta + \cdots \Bigr] e^{2(\varphi - \varphi_\star)}\,,
\ee
where the ratio between $a \simeq \sum_\varv \delta T_\varv = 2 \, \delta T_{\rm avg}$ and $b \simeq 2g_\ssR \mu/\kappa^2$ is again a dial we can exploit. Assuming $0 < \eta < 1$ the extrema are determined by the vanishing of
\be \label{eq:Uprimequad}
 U' \simeq \left[ 2a + 2b (\varphi - \varphi_c)^\eta + \frac{\eta b}{ (\varphi-\varphi_c)^{1-\eta}} + \cdots \right]  e^{2(\varphi - \varphi_\star)} \,,
\ee
which has solutions in the regime $|\varphi - \varphi_c| \gg 1$ of the form
\be
 \varphi_\star - \varphi_c \simeq  \left(- \frac{a}{b} \right)^{1/\eta} \,,
\ee
which for small $\eta$ is true even if $a/b = \kappa^2 \delta T_{\rm avg}/g_\ssR \mu$ is only moderately large and negative. (For instance choosing $\eta = \frac13$ and $\kappa^2 \delta T_{\rm avg} \sim - 4 g_\ssR \mu$ gives $\varphi_\star - \varphi_c \simeq - 64$.) At this point we have
\be
 U_\star = a + b (\varphi_\star - \varphi_c)^\eta \simeq - \frac{\eta b}{ 2(\varphi_\star - \varphi_c)^{1-\eta}} \simeq - \frac{\eta b}{2} \left( -\frac{b}{a} \right)^{(1-\eta)/\eta}  \,.
\ee
Small $\eta$ has the virtue of amplifying both the size of $\varphi_\star$ and the suppression of $U_\star$, although not in a way that seems phenomenologically viable for both at the same time.

\subsubsection*{Case IV: Exponential $\xi$}

Next consider an example whose solutions are perturbatively close to the asymptotic runaway. This example is similar to the scaling case examined for the UV vortex completion in \cite{Companion2}, where $T = T_0 + \delta T$ and $\xi = \xi_0 + \mu \, e^{s \varphi}$, where $g_\ssR \xi_0 = - \kappa^2 T_0$ and our main interest is in $s$ not far from zero. In this case $y = (g_\ssR\mu/2\pi) \, e^{s\varphi}$ so $y' = sy$, and
\be
 U = \Bigl( a + b\, e^{s\varphi} + \cdots \Bigr) e^{2(\varphi- \varphi_\star)}  \,,
\ee
with $a \simeq 2 \delta T$ and $b \simeq 2g_\ssR\mu/\kappa^2$. Then the non-runaway solutions to $U'=0$ satisfy
\be
 2a + b(2+s) e^{s\varphi_\star} + \cdots = 0 \,.
\ee
If $e^{s\varphi_\star}$ is small enough to drop all but the first two terms we have
\be \label{eq:sstar}
 e^{s\varphi_\star} \simeq -\frac{2a}{b(2+s)} \simeq -\frac{2 \kappa^2 \delta T}{(2+s) g_\ssR \mu} \,,
\ee
which requires $a$ and $b$ to have opposite signs. The value of $U$ at the extremum is
\be \label{eq:Ustars}
 U_\star \simeq - \frac{s b}4 \; e^{s\varphi_\star} \simeq \frac{sa}{2(2+s)} \simeq \frac{s \delta T}{2+s}  \,.
\ee
The factor of $s$ found in $U_\star$ can be understood because when $s \to 0$ the potential becomes scale-invariant and so must then be minimized at $U_\star = 0$ with $\varphi_\star \to - \infty$.

If $s > 0$ then having small $e^{\varphi_\star}$ means we must also have $|a| \ll |b|$ (which corresponds to $\kappa^2 |\delta T| \ll |g_\ssR\mu| \ll 1$). In this case $U$ asymptotes to zero as $\varphi \to - \infty$ from below (above) if $a$ is negative (positive), so the extremum is a minimum if $a < 0$ and $b > 0$ ({\em ie} when $g_\ssR \mu > 0$ and $\delta T < 0$) in which case $U_\star < 0$.

Conversely, if $s < 0$ then having small $e^{\varphi_\star}$ means we instead must have $|a| \gg |b|$ (and so $|g_\ssR\mu| \ll \kappa^2 |\delta T| \ll 1$), and in this case it is for $b < 0$ and $a > 0$ ({\em ie} for $g_\ssR \mu < 0$ and $\delta T > 0$) that the above root is a minimum. Writing $s = -\sigma$
\be
 e^{\sigma\varphi_\star} \simeq -\frac{b(2-\sigma)}{2a} \simeq -\frac{g_\ssR \mu (2-\sigma)}{2 \kappa^2 \delta T} \,,
\ee
which again requires $\mu$ and $\delta T$ to have opposite signs. The value of $U$ at the extremum is
\be
 U_\star \simeq \frac{\sigma b}4 \; e^{-\sigma\varphi_\star} \simeq -\frac{\sigma a}{2(2-\sigma)}  \simeq -\frac{\sigma \delta T}{2-\sigma}  \,,
\ee
which is again negative and order $\sigma \delta T$. To be much smaller than $\delta T$ we would need $\sigma \ll 1$.

The corresponding physical KK scale is
\bea \label{ellvsrB}
 \ell = r_\ssB \, e^{-\varphi_\star/2} &\sim& \left( \frac{\kappa}{2g_\ssR} \right) \left( -\frac{g_\ssR \mu}{2 \kappa^2 \delta T } \right)^{1/2s} \qquad (\hbox{if $s > 0$}) \\
 &\sim& \left( \frac{\kappa}{2g_\ssR} \right) \left( - \frac{2 \kappa^2 \delta T }{g_\ssR \mu}   \right)^{1/2\sigma} \qquad (\hbox{if $s = -\sigma < 0$}) \,,\nn
\eea
and the classical prediction for the mass of the would-be zero mode is
\be
 m_\varphi^2 = \frac{1}{2} \kappa_4^2 U^{\prime \prime}_\star \sim - s \kappa_4^2 \delta T  \simeq - (2+s) \kappa_4^2 U_\star \,.
\ee

The minimum found above is most interesting when $|s| \ll 1$, for two reasons.
First, small $s$ ensures that $e^{\varphi_\star}$ can be {\em extremely} small even if both $\kappa^2 \delta T$, $g_\ssR \mu$ and their ratio are only moderately small. For example, taking $\kappa^2 \delta T \sim 0.3$ and $g_\ssR \mu \sim - 0.0003$ gives $\kappa^2 \delta T/g_\ssR \mu \sim -10^{3}$ and so $s = - \sigma \sim -0.1$ gives the enormous hierarchy $e^{\varphi_\star} \simeq r_\ssB/\ell \sim 10^{-15}$ appropriate to a picture with micron-sized extra dimensions \cite{SLED, ADD} when the bulk is controlled by TeV scale physics. Such large radii arise because the choice $0 < |s| \ll 1$ makes the setup close to scale-invariant, and so the potential in this limit is close to its runaway form, $U \sim U_0 e^{2\varphi}$. The small scale-breaking parameters then give a weak $\varphi$-dependence to the prefactor $U_0$, creating a minimum out at large negative $\varphi$. The minimum occurs at large $-\varphi$ precisely because of the potential's close-to-runaway form.

Small $|s|$ is also interesting because of the suppression implied by \pref{eq:Ustars} for the value of $U_\star$. As mentioned earlier, this suppression arises generically because the system becomes classically scale invariant in the $s \to 0$ limit, and so $U_\star$ must vanish in this limit. Effectively this converts Weinberg's runaway no-go from a bug to a feature, with weak scale-breaking driving $U_\star$ to be small precisely because the minimum gets driven out to infinity in the scale-invariant limit. As before, however, although both large $\ell$ and small $U_\star$ are possible, no one choice of parameters gets both right at the same time (without very precise tuning to make $|a/b|$ extremely close to unity.

\subsubsection*{When large $\varphi$ does not imply large dimensions}

Equating large negative $\varphi_\star$ to a large hierarchy between KK size, $\ell$, and brane size, $r_\ssV$, (as done in the previous examples) implicitly makes an assumption about the $\varphi$-dependence of $r_\ssV$. The issue is whether or not obtaining large $e^{-\varphi_\star}$ --- {\em eg} \pref{ellvsrB} --- is sufficient to imply a large hierarchy between $\ell$ and the transverse brane size, $r_\ssV$. It need not be, depending on the other microscopic details that determine $r_\ssV$. In particular it depends on how the brane size itself depends on $\varphi_\star$.

For instance, the UV completion considered in \cite{Companion2} provides an example where the connection between large $\varphi_\star$ and large $\ell/r_\ssV$ can fail. In this example the branes are resolved in the UV as Nielsen-Olesen vortices \cite{NOSolns} with tension, $T_\varv \simeq v^2$, set by a scalar vev, $v$, and brane-localized flux, $\xi \simeq (2\pi n \varepsilon/e) e^{s \varphi}$, set by a dimensionless mixing parameter, $\varepsilon$, an integer, $n$, and a gauge coupling, $e$. In this UV completion the physical size of the vortex is $r_\ssV^{-1} \simeq v \hat e(\varphi) = e v \exp \left[ \frac12 (1+2s) \varphi \right]$, which turns out to inherit a dependence on $\varphi$ from the effective coupling $\hat e(\varphi)$. Consequently $r_\ssV/\ell$ can be related to the tension and localized flux by
\be \label{eq:lrzeta}
 \frac{r_\ssV}{\ell} = \left( \frac{2g_\ssR}{\kappa} \, e^{\varphi_\star/2} \right) \left( \frac{1}{ev} \, e^{-(1+2s)\varphi_\star/2} \right) = \left( \frac{2g_\ssR}{e \kappa v} \right) e^{s \varphi} = \frac{g _\ssR \, \xi}{n  \pi \varepsilon \kappa v}  = \frac{g _\ssR \, \xi}{n  \pi \varepsilon \sqrt{\kappa^2 T}} \,,
\ee
for any choice of $s$ or $\varphi$. What is important here is that $r_\ssV/\ell$ is $\varphi$-independent when expressed in terms of the parameters, $T$ and $\zeta$, appearing in the brane effective lagrangian, since these are the combinations that are relevant to the long-distance physics governing the size of $\ell$. As a result, in this particular model it doesn't matter how large $\varphi_\star$ is when predicting $r_\ssV/\ell$.

Notice that this line reasoning relies on {\em all} of $\xi$ depending on $\varphi$ in the same way, rather than there being several contributions involving different scales and depending differently on $\varphi$. This is why it does not also apply to the previous examples, for which $\xi = \xi_0 + \delta \xi(\varphi)$.

\subsection{Scenarios of scale}

Before turning to the robustness of the above examples it is useful to have some idea in mind for the the mass scales appearing in all sectors of the theory. This is important when estimating quantum corrections in particular, since for naturalness problems the heaviest scales are usually the most dangerous. We also imagine at least one brane lagrangian being modified to include brane-localized particles, including the known Standard Model (SM) particles.

There are several mass scales potentially in play: the inverse brane width, $M \sim 1/r_\ssV$; the SM electroweak scale, $m$; and the scale set by bulk couplings, $\kappa^{-1/2}$ and $g_\ssR^{-1}$. Without loss we may shift $\varphi$ in the bulk so that $\varphi = 0$ corresponds to $g_{\ssR}^{-1} \sim \kappa^{-1/2} \sim M_g$ defining the same scale. The effective bulk gauge coupling, $g_\star = g_\ssR \, e^{\varphi_\star/2 }$ and the KK scale, $m_\KK \sim 1/\ell \sim g_\star/\kappa = (g_\ssR/\kappa) e^{\varphi_\star/2}$, are then computed from these once the dilaton is stabilized at $\varphi = \varphi_\star$. We assume the hierarchy
\be
 M_g \sim \kappa^{-1/2} \sim g_\ssR^{-1} \gg M \gg m \,,
\ee
and ask how loops might depend on these scales.

It is also useful to imagine the UV completion of the brane eventually becomes supersymmetric at high enough energies, since this is likely necessary to deal with naturalness at the highest scales possible. This could happen at the string scale if the brane UV completes as an object within string theory, or it could happen above or below the scale $M$ if the branes UV complete as vortices in a higher-dimensional field theory. For concreteness we consider the vortex completion, since the extension to string theory of the system used here remains an open question \cite{6DStrings}. Since our goal is to explore extra-dimensional approaches to the hierarchy problem, we always take the brane SUSY-breaking scale, $M_s$, much larger than electroweak scales: $M_s \gg m$.

If we choose $M_s \ll M$ then the vortex sector would be supersymmetric (in that it would preserve at most half of the supersymmetries of the bulk \cite{PolHughes}) with the branes likely arising as BPS solutions. Until distorted by supersymmetry-breaking effects (if any) we would then expect the largest contributions to $T$ and $\zeta$ to be $\phi$-independent, with $T = T_s \sim \cO(M^4)$. The supersymmetry breaking such branes generically imply for the bulk sector is then minimized if the branes carry the supersymmetric amount of flux \cite{AccidentalSUSY}, so we take $\kappa^2 T_s = \pm \frac12 \, g_\ssR \zeta_s$. This implies $\zeta_s \sim \kappa^2 T_s/g_\ssR \sim M^4/M_g^3 \ll M$ in magnitude. These assumptions ensure a flat potential, $U = 0$, for $\varphi$ and allows supersymmetry to protect this shape from scales higher than $M_s$, leaving nontrivial corrections to the low-energy theory (where we can try to estimate them).

We expect nonzero $\delta T = T(\varphi) - T_s$ and $\delta \zeta = \zeta(\varphi) - \zeta_s$ once effects of the SUSY-breaking brane sector are included. This includes but need not be limited to the SM sector (which is assumed to be localized to one of the branes/vortices). On dimensional grounds, if SUSY breaks on the branes with scale $M_s$ such that $m \ll M_s \ll M$ we expect the dominant deviations from the supersymmetric limit to be of order $\delta T(\varphi) \sim M_s^4$ and $\delta \zeta(\varphi) \sim M_s$. If the supersymmetry breaking physics respects the bulk scale invariance then $\delta T$ and $\delta \zeta$ remain $\varphi$-independent; otherwise not.

Suppose the supersymmetry-breaking sector {\em does} break scale invariance but only through the localized flux term as examined above, so $T = T_s + \delta T$ with $\delta T \sim M_s^4$ and $\zeta = \zeta_s + \delta \zeta$ with $\delta \zeta(\varphi) \sim M_s \, f(\varphi)$, for some function $f(\varphi)$, although the precise form for $f$ is not yet crucial. Assuming $M_s \gg M^4/M_g^3$ then there should exist a value, $\varphi = \varphi_s$, for which $U(\varphi_s) = 0$ because $\zeta(\varphi_s)$ accidentally takes the supersymmetric value corresponding to $T = T_s + \delta T$,
\be
 \pm \frac12 \, g_\ssR \zeta(\varphi_s) = \pm \frac12 \, g_\ssR \left[ \zeta_s + \delta \zeta(\varphi_s) \right] \sim \kappa^2 T = \kappa^2 (T_s + \delta T) \,.
\ee
We imagine the value, $\varphi_s$, where this occurs to be moderately large (of order -75 or so in the extreme case of very large dimensions).

This scenario fits very cleanly into the class of models for which the perturbative methods explored earlier apply, with $y(\varphi) \sim g_\ssR \delta \zeta (\varphi) \sim g_\ssR M_s \delta f(\varphi) := g_\ssR M_s [f(\varphi) - f(\varphi_s)] = \cO(M_s/M_g)$. If $\delta f$ varies slowly enough to be approximated as linear near $\varphi_s$ the analysis of earlier sections would predict a minimum with $\varphi_\star - \varphi_s \simeq - \frac12$ at which point the classical 4D energy density is $U_\star \sim - g_\ssR M_s/\kappa^2$. Other forms for $f(\varphi)$ would predict different scalings.

Finally, loops of Standard Model particles should also contribute to $T$ and $\zeta$ and further perturb them away from their supersymmetric relationship, by an amount at least $\delta T_\SM \sim m^4$ and $\delta \zeta_\SM \sim \epsilon \, m$ (where $\epsilon \lsim 1$ is a dimensionless measure of the strength with which the SM sector couples to the bulk gauge field). Even if not supersymmetric, such SM contributions need not contribute any $\varphi$-dependence if they preserve scale invariance.

There are two natural ranges of values to think through, depending on whether our interest is in the electroweak hierarchy (quantum corrections to scalar masses) or the cosmological constant problem (quantum corrections to vacuum energies). We consider each of these briefly in turn.

\medskip\noindent{\em Electroweak Hierarchy}

\medskip\noindent For applications to the electroweak hierarchy we ask the extra dimensions to be large and take the large scales all to be of order the electroweak scale, with the minimal hierarchy required for control of approximations. In this case the premium is on predicting the value of $\varphi_\star$ from first principles to ensure sufficiently large $\ell/r_\ssB$ using only a relatively modest hierarchy amongst lagrangian parameters, and we are happy to fine-tune away any cosmological constant. This can be done, for example, if the vortex size, $r_\ssV$, is $\varphi$-independent and controls the supersymmetric brane physics at scale $M$, and the supersymmetry-breaking brane physics at scale $M_s$ generates an exponential $\delta \zeta \sim M_s \, e^{s\varphi}$.

Taking for illustrative purposes $M_g \sim 50$ TeV, $M \sim M_s \sim 5$ TeV and $m \sim 100$ GeV with $s \simeq 0.2$ then gives $M_g \ell \sim 10^{15}$, which is in the ballpark required. Such a dynamical explanation for the exponentially large size of $\ell$ elevates the large-dimensional models \cite{ADD} to a footing similar to their warped competitors \cite{RS}, although this would be more satisfying with a more explicit picture for the SUSY-breaking brane physics to see more explicitly how it generates the required $\varphi$-dependence for $\zeta$ and $T$.

The challenge and opportunity in this scenario is to better construct the SUSY breaking physics, partly to see what signals it could imply at the LHC. There is clearly some freedom to dial scales somewhat, though if $M_s$ and $M_g$ are both taken much larger than the electroweak scale we must again ask what protects the value of the Higgs mass on the brane. Implicit in any such model is that whatever quantum gravity eventually kicks in at $M_g$ does not allow the higher scales to feed into the Higgs mass and thereby ruin the naturalness of the low-energy picture.

\medskip\noindent{\em Vacuum Energies}

\medskip\noindent Although the ideal situation would be to explain the observed Dark Energy density, it would already be progress on the cosmological constant problem to suppress $U_\star$ below the electroweak scale. This requires the classical contribution be smaller than the known quantum effects (usually not hard), while choosing parameters so that the quantum effects themselves can be smaller than the electroweak scale (usually much harder). The hope here is that because SM loops generate changes to the brane tension, $\delta T \sim m^4$, we seek choices that keep this from directly contributing to $U_\star$.

A best case in this type of scenario is to imagine that all physics couples to $\varphi$ in the scale-invariant way down to as low an energy (say $\mu$) as possible. If $\mu \ll m \ll M_s \ll M$ then this implies the UV physics is to first approximation scale invariant though not supersymmetric, so that $T$ and $\zeta$ are constants for which $\kappa^2 T$ and $g_\ssR \zeta$ are not similar in size.

In this case we imagine the scale-invariance breaking at scale $\mu$ introduces a $\varphi$-dependence only to $\delta\zeta$, in such a way that $\zeta$ accidentally passes through the supersymmetric point, $\zeta \sim \pm 2\kappa^2 T/g_\ssR$ at $\varphi_s \sim - 75$ or so. This ensures the extra dimensions can be very large (best of all would be in the micron range) as desired. Provided the variation in $\varphi$ is slow enough to justify Case I above, the classical prediction for $U_\star$ is negative\footnote{Having $U_\star < 0$ need not be a problem if its magnitude is small enough that the vacuum energy is dominated by its quantum parts (which must then be positive).} with magnitude $\sim g_\ssR \mu/\kappa^2$. Choosing $M_g$ as low as possible (in the 10 TeV regime, say) then gives a suppression of $U_\star$ relative the electroweak scale by of order $g_\ssR \mu$.

How much suppression depends on how small $\mu$ can be, which requires a better theory of the origins of the $\varphi$-dependence. Since $U_\star \sim \mu M_g^3$ we see that having $|U_\star| \lsim (10^{-2} \; \hbox{eV})^4$ and $M_g \sim 10$ TeV requires fantastically small values like $\mu \lsim |U_\star|/M_g^3 \sim 10^{-47}$ eV. To the extent that useful progress on lowering $U_\star$ below the electroweak scale requires scale-invariant couplings of $\varphi$ to ordinary matter, the obstacle is likely to be solar-system constraints on the existence of light Brans-Dicke scalars with gravitational couplings.

\subsection{Robustness}

As for any approach to naturalness problems the key question concerns robustness of the result. One must check whether conclusions survive the inclusion of subdominant terms in the various approximations being made. Although a full analysis of all of these corrections goes beyond the scope of this article, we make a few preliminary estimates of the size of some of the usual suspects.

\subsubsection*{Potentially fragile choices}

Assessments of robustness turn on the generality of the choices for parameters in the classical theory. Because it is the branes that are responsible for breaking supersymmetry we might expect that it is choices made for the brane actions in particular that are the most susceptible to perturbations (such as by receiving quantum corrections once these are included).

The basic choices used in previous sections concern the magnitude and $\phi$-dependence of the brane action, parameterized by the small dimensionless quantities $\kappa^2 T(\phi)$ and $g_\ssR \zeta(\phi)$ for each of the branes. In particular the previous sections make two non-generic assumptions about the brane action:
\begin{itemize}
\item We choose no $\phi$-dependence for $T$ but allow $\phi$-dependence for $\zeta$;
\item We dial freely the relative magnitudes of $\kappa^2 T$ and $g_\ssR \zeta$.
\end{itemize}
It is the sensitivity of these choices to quantum corrections on which we focus.

\subsubsection*{Some quantum estimates}

UV sensitive quantum corrections in this type of model come in two broad classes: quantum corrections to the bulk lagrangian due to loops of bulk fields; and quantum corrections to the brane lagrangians due to loops of fields on the brane and loops involving bulk fields located close to the brane. In both cases it is loops of the most massive particles that are potentially the most dangerous.

\medskip\noindent{\em Corrections to the Bulk Sector}

\medskip\noindent
Loops within the supergravity describing the bulk have been studied in some detail \cite{AccidentalSUSY,DistributedSUSY,UVsens}, and although loops of individual massive states do renormalize all terms in the bulk and brane lagrangians their contributions to the bulk lagrangian tend to cancel once summed over 6D supermultiplets \cite{UVsens}. The only bulk renormalizations that survive these cancelations are renormalizations of those interactions allowed by bulk supersymmetry, for which we do not make any special requirements.

This is required physically because UV modes far from the branes effectively do not know that supersymmetry is broken. The UV dangerous renormalizations coming from the supersymmetric sector are those that renormalize the non-supersymmetric brane physics. These should not be dangerous to the extent we do not make special assumptions about the sizes (or the dependence on bulk fields) of couplings like $\check T$ and $\zeta$ in the brane action.

From the point of view of the vacuum energy, the most dangerous renormalizations of the bulk are dimension-four interactions involving curvature squared terms (and their partners under supersymmetry) since these can acquire renormalizations proportional to the squared-mass, $\cM^2$, of the massive bulk supermultiplet \cite{AccidentalSUSY,DistributedSUSY,UVsens}. These can generate contributions to the 4D vacuum energy of order $\cM^2/\ell^2$, and so be larger than the $1/\ell^4$ desired to describe Dark Energy in SLED models. But they are generically smaller than the $\cO(\cM^4)$ contributions described below, and so represent a lesser worry than the brane renormalizations we describe next.

\medskip\noindent{\em The Brane Sector: Bulk Loops}

\medskip\noindent
Loops of bulk fields involving virtual particles physically near the branes also renormalize the brane lagrangian, as computed in \cite{AccidentalSUSY, DistributedSUSY}. These loops turn out not to be dangerous for our two brane choices, however, for two reasons.

The first statement is that although bulk loops contribute of order $\cM^4$ to the brane tension, they do not introduce nontrivial $\phi$-dependence to the tension if this was not already present because of the underlying scale invariance of the bulk system. Secondly, bulk loops involving massive multiplets that carry gauge charge can also renormalize $\zeta$. But because the correction is of order $\delta \zeta \sim g_\ssR^2 \cM^2 \zeta$ \cite{AccidentalSUSY} it is technically natural (from the point of view of these loops) to choose $\zeta$ to be small.

\medskip\noindent{\em The Brane Sector: Brane Loops}

\medskip\noindent
Massive fields localized on the branes are among the most dangerous (and arguably the most difficult to understand) from the point of view of naturalness, because these fields can be heavy and are not constrained by supersymmetry (at least at scales below $M_s$).  In principle these include loops of familiar SM fields that are the origin of the cosmological constant problem in the first place.

Integrating out such particles of mass $M$ generically renormalizes the brane tension by an amount of order $M^4$, so we run into naturality problems as soon as we must demand $\delta T$ be smaller than this. For applications to the cosmological constant problem this is why all contributions to $U_\star$ of order $\delta T$ are not regarded as being progress.

In general such loop contributions to $T$ could also play a role by introducing nontrivial $\phi$-dependence, although this can be protected against by demanding the couplings of the brane matter to preserve scale invariance. For SM fields this is trouble to the extent that it gives them Brans-Dicke couplings \cite{BDS} to the light scalar $\varphi$ of gravitational strength \cite{GravCoup}, which are ruled out phenomenologically (for sufficiently light $\varphi$) by PPN solar-system tests of gravity \cite{BDBounds}. Of course, mechanisms exist for weakening the couplings of light scalars \cite{ABS, Chameleon}, usually by making these couplings $\varphi$- or environment-dependent or by making the scalar massive enough not to mediate a sufficiently long-range force. Although much model-building could be forgiven if progress could be made on the cosmological constant problem, we regard this to be a real worry whose resolution goes beyond the scope of this (already very long) paper.

The same kinds of problems need not be a worry for brane corrections to $\zeta$, however, because these cannot be generated unless the field in the loop already couples to the bulk gauge field. Brane-generated contributions to $\delta \zeta$ should be easy to suppress simply by not coupling heavy brane particles to this field.

\section{Discussion}
\label{section:discussion}

This paper's aim is to carefully determine how codimension-two objects in 6D supergravity back-react on their in environment through their interactions with the bulk metric, Maxwell field and dilaton, and how this back-reaction gets encoded into the effective potential of the low-energy 4D world below the KK scale.

To this end, we construct the corresponding four-dimensional effective theory, and show how the flux quantization conditions of the UV theory are brought to 4D by a four-form gauge flux dual to the Maxwell field. The 4D theory generically contains a light scalar dilaton to the extent that the branes do not strongly break the classical bulk scale invariance. We identify the scalar potential for this scalar and show at the linearized level that it agrees with what is obtained by explicitly linearizing the higher-dimensional field equations. This calculation in particular corrects some errors in \cite{BLFFluxQ}, which misidentified some of the boundary conditions associated with the brane-localized flux term.

We confirm the result of \cite{BLFFluxQ} that the breaking of scale invariance by the branes can lead to modulus stabilization and allow explicit computation of the extra-dimensional size, in a codimension-two version of the Goldberger-Wise mechanism \cite{GoldWis}. We confirm that this size can be exponentially large in the brane couplings. A moderate hierarchy of order $75$ amongst the brane couplings can be amplified to produce enormous extra dimensions in this way, thereby fixing a long-standing problem with the use of large extra dimensions to solve the electroweak hierarchy problem.

For the particular choice of near scale-invariant couplings we can (but need not) also find some parametric suppression in the value of the on-brane curvature and dilaton mass, although for those examined so far this suppression seems fairly weak. We are unable to find simple examples which both generate exponentially large dimensions and suppress the classical vacuum energy (though we also are unable to prove this to be impossible).

Although we make preliminary estimates about the size of quantum corrections and the robustness of the parametric suppressions of the potential, we leave a more detailed treatment to later work.

\section*{Acknowledgements}

We acknowledge Florian Niedermann and Robert Schneider for collaborations at early stages of this work as well as discussions about this paper and \cite{Companion, Companion2} while they were in preparation. We thank Ana Achucarro, Asimina Arvanitaki, Savas Dimopoulos, Gregory Gabadadze, Ruth Gregory, Mark Hindmarsh, Stefan Hoffmann, Leo van Nierop, Massimo Porrati, Fernando Quevedo and Itay Yavin for useful discussions about self-tuning and UV issues associated with vortices and brane-localized flux. The Abdus Salam International Centre for Theoretical Physics (ICTP), the Aspen Center for Physics, the Kavli Institute for Theoretical Physics (KITP), the Ludwig-Maximilian Universit\"at, Max-Planck Institute Garsching and the NYU Center for Cosmology and Particle Physics (CCPP) kindly supported and hosted various combinations of us while part of this work was done. This research was supported in part by funds from the Natural Sciences and Engineering Research Council (NSERC) of Canada, and by a postdoctoral fellowship from the National Science Foundation of Belgium (FWO), by the Belgian Federal Science Policy Office through the Inter-University Attraction Pole P7/37, the European Science Foundation through the Holograv Network, and the COST Action MP1210 `The String Theory Universe'. Research at the Perimeter Institute is supported in part by the Government of Canada through Industry Canada, and by the Province of Ontario through the Ministry of Research and Information (MRI). Work at KITP was supported in part by the National Science Foundation under Grant No. NSF PHY11-25915. Work at Aspen was supported in part by National Science Foundation Grant No. PHYS-1066293 and the hospitality of the Aspen Center for Physics.

\appendix

\section{Scale invariant solutions}
\label{app:SIsolns}

In this appendix we present the details of well-known solutions that exist when the branes are scale invariant. We first describe the Salam-Sezgin solution \cite{SS} that applies when there are no branes, and we then show how this solution generalizes to the rugby ball solution \cite{SLED} in the case where the branes are identical, scale invariant, and supersymmetric.

\subsection{Salam-Sezgin solution}
\label{app:ss}

In the absence of branes, it is consistent to assume a trivial warp factor $W =1$ and no dilaton profile $\phi^\prime = 0$. The second of these conditions is satisfied as long as the source terms in the dilaton field equation vanish
\be \label{AppAphi}
 \Box \phi = e^{\phi} \left[ \left( \frac{2 g_\ssR^2}{\kappa^4} \right) - \frac{1}{2} Q^2 \right] = 0 \,,
\ee
where we eliminate the bulk field strength in terms of $Q$ using $A_{\rho \theta} = Q B e^{\phi}$. The constant $Q$ is fixed below by flux quantization and for generic values of $Q$ the above equation is only solved by taking the runaway solution: $\phi \to - \infty$. The exception is if flux quantization returns the specific Salam-Sezgin value, $Q = \pm Q_{s} := \pm 2 g_\ssR / \kappa^2$, in which case \pref{AppAphi} is solved for any constant: $\phi = \varphi$.

With this choice of $Q$, the bulk metric function satisfies the field equation
\be \label{eq:Rrb}
 \frac{B''}{B} = - \frac{ e^{\varphi} }{ L_{s}^2} \,,
\ee
where $L_{s}  = r_\ssB = \kappa / 2 g_\ssR$. The solution is
\be
 B_{s} = r_\ssB e^{-\varphi/2} \sin(\rho \, e^{\varphi/2} / r_\ssB ) \,,
\ee
and we conclude the extra dimensions are spherical with proper radius $\ell_{s}^2 = r_\ssB^2 e^{ - \varphi}$.

Consistency requires verifying the flux-quantization condition returns $Q = Q_s$. To check we evaluate
\be \label{eq:appendixflux}
 \frac{N}{g_\ssA} = \int_0^{\ell_s} \d \rho \, A_{\rho \theta} = Q  \int_0^{\ell_s} \d \rho B e^{\phi} =  \left( \frac{Q}{Q_s} \right) \frac{1}{g_\ssR} \,,
\ee
where $g_\ssA$ is the gauge coupling of the background gauge field (which in principle could differ from $g_\ssR$ if this field gauges a group other than the $R$-symmetry for which $g_\ssR$ is the coupling). We see that only the supersymmetric choices $g_\ssA = g_\ssR$ and $Q/Q_s = N = \pm 1$ are consistent with $\phi = \varphi$ being a finite constant, and because $Q = \pm Q_s$ the value, $\varphi$, remains undetermined by the field equations.

\subsection{Supersymmetric rugby ball}
\label{app:rugbysoln}

Many of the nice properties of the Salam-Sezgin solution are preserved if identical, scale invariant, supersymmetric branes are added to the system, with action
\be
 S_{\rm branes} = -\sum_\varv \int_{\Sigma_\varv(u)} \d^4 u \sqrt{-\gamma} \left( T - \frac{1}{ 4!} \zeta  \, \varepsilon^{\mu\nu\lambda\rho} F_{\mu\nu\lambda\rho} \right) \,,
\ee
where $\Sigma_\varv(u)$ denotes the worldsheet of each brane, parameterized by the four coordinates $u^\mu$. By assumption $T$ and $\zeta$ are the same for both branes and independent of the dilaton (as required for the branes not to break the classical bulk scale invariance). These choices are necessary if the branes are not to source gradients of the warp factor or dilaton, making it still consistent to assume $W =1$ and $\phi^\prime = 0$.

As before, the condition of constant $\phi$ requires bulk sources in the dilaton field equation to vanish, and so flux quantization must return the same value for $Q$ as in the Salam-Sezgin solution: $\bar Q = \pm Q_{s} = \pm 2 g_\ssR / \kappa^2$. This choice of $Q$ also preserves the radius of the extra dimensions so the rugby ball metric function is solved by
\be
 B = \alpha L_s e^{-\varphi/2} \sin(\rho\, e^{\varphi/2} / L_s) \,.
\ee
Note the presence of the constant $\alpha$ in this solution, which physically represents a conical singularity at the poles of the sphere with defect angle $\delta = 2\pi(1 - \alpha)$. This differs from the Salam-Sezgin value, $\alpha_{s} = 1$, because the presence of branes modifies the boundary condition of the bulk metric function $B$ at the position of the branes \pref{eq:defect} to satisfy
\be
  B'(\rho_\varv) = \alpha = 1 - \frac{\kappa^2 T}{2 \pi} \,,
\ee
where the sign assumes the derivative is in the direction away from the brane and $T$ is the brane's tension. Nonzero defect angles make the bulk resemble a rugby ball rather than a sphere.

The other effect of the branes is to introduce a localized piece of $A_{\rho \theta}$ at the brane positions, and this modifies the flux-quantization condition \pref{eq:appendixflux} to become
\be
  \frac{N}{g_\ssA} = \int_0^{\ell_s} \d \rho \, A_{\rho \theta} = Q  \int \d \rho \, B e^{ \phi} - \frac{1}{2 \pi} \sum_\varv \zeta \,,
\ee
showing that $\zeta/2\pi$ describes that amount of the total gauge flux that is localized in this way. Evaluating as before gives
\be
 \frac{N}{g_\ssA} = \frac{\alpha }{g_\ssR}\left( \frac{Q}{Q_s} \right) - \frac{1}{2 \pi} \sum_\varv \zeta = \frac{1}{g_\ssR} \left( \frac{Q}{Q_s} \right) \left(1  - \frac{ \kappa^2  T }{2 \pi} \right) - \frac{1}{2 \pi} \sum_\varv \zeta \, ,
\ee
which shows how the brane-localized flux compensates for the reduction of bulk volume caused by the defect angles.

Flux quantization is only consistent with constant $\phi$ if it returns $Q = \pm Q_s.$ Having source branes can allow this if $T$ and $\zeta$ are related by
\be \label{eq:branesusy}
 \kappa^2 T = \mp g_\ssR \sum_\varv \zeta = \mp 2 g_\ssR \zeta  \,,
\ee
in addition to the bulk conditions $g_\ssA = g_\ssR$ and $N= \pm 1$. This brane condition also turns out to be required by demanding supersymmetry not be broken by the presence of the branes \cite{AccidentalSUSY}, showing how supersymmetry again ensures the value $Q = Q_{s}$ required for a flat potential that does not determine the value $\phi = \varphi$.

\section{Linearized solutions}
\label{app:Linearizations}

We now assume that the branes are perturbatively close to the identical, scale-invariant supersymmetric ones just described. However, the perturbations we consider to the tension and localized flux need not respect scale invariance and can differ at each brane
\be
 \zeta_\varv = \zeta_0 + \delta \zeta_\varv(\phi) \qquad \text{and} \qquad  T_\varv = T_0 + \delta T_\varv(\phi) \,,
\ee
where $T_0$ and $\zeta_0$ satisfy \pref{eq:branesusy}.

We track the effects of these perturbations on the the bulk fields by solving the entire set of field equations, including the equations for warping, the dilaton profile, and flux quantization, at linear order in the perturbations. When the brane perturbations break scale invariance, we also solve for the stabilized value of the zero mode, $\varphi = \varphi_\star$, to linear order. We also calculate the 4D effective potential for $\varphi$ at the linearized level, and show how it reproduces this stabilized value of the zero mode computed with the full 6D theory.

\subsubsection*{Full field equations}

We first present the set of field equations and boundary conditions to be solved.

Because of the scale invariance of the unperturbed theory it is useful to switch to the following scale invariant variables
\be
 b := e^{\phi/2} B \qquad \hbox{and} \qquad \d \sigma = e^{\phi/2} \, \d \rho \,.
\ee
With these the undifferentiated dilaton only appears in the field equations through scale-breaking terms. Since these terms are by assumption perturbatively small, we can simply replace the dilaton factor $\phi$ appearing there with the zero mode $\varphi$. These variables also simplify the linearization of the scale invariant terms in the field equations since the background dilaton solution reads $\bar \phi^\prime = 0$ (where bars denote background quantities). Additionally, the equations simplify if we rewrite the warp factor as
\be
 W^4 = e^\omega \,,
\ee
so we can perturb around the background solution $\bar \omega = 0$. In these new variables the background of the bulk metric function simplifies to $\bar b = \bar \alpha \bar L \sin(\sigma / \bar L )$, with $\bar\alpha$ determined by $\kappa^2 T_0$ and $\bar L = L_s = r_\ssB$.

Since our interest is in computing the shape of the zero-mode potential we also follow Refs.~\cite{BLFFluxQ} and add a stabilizing current to the bulk action
\be
 \Delta S_\ssJ  = - \int \d^6 x \sqrt{-g} \,  J = -\int \d^4 x \int \d \theta \int \d \sigma  \,  J \, b \, e^{\omega - \phi} \,.
\ee
Choosing $J$ appropriately allows us to investigate values of $\varphi$ away from the minimum of the potential while still solving all of the field equations. In particular, we read the equation that would have determined the stabilized value $\varphi = \varphi_\star$ as instead to be solved for $J(\varphi)$, allowing us to trace the shape of the effective potential for $\varphi$. Then, $\varphi = \varphi_\star$ corresponds to $J = 0$.

To solve for the perturbations to the bulk metric function and warp factor, we desire two linear combinations of the Einstein equations \pref{BavR4-v1} - \pref{BnewEinstein} that contain second derivatives of the metric fields, and no factors of the 4D curvature. The first of these reads
\be
 \left[ e^\omega \left( b' - \frac12 \, b \, \phi' \right) \right]'  = - \kappa^2  b \, e^\omega  \left( \frac{3 Q^2}{4 } \, e^{-2\omega} + \frac{g_\ssR^2}{\kappa^4}  + \frac{1}{2} J e^{-\phi} \right) \,,
\ee
and is to be solved for $b$. (From here on primes on bulk fields denote differentiation with respect to $\sigma$ rather than $\rho$.)

The other relevant Einstein equation is
\be
  \omega'' + \omega' \, \phi'  + \frac{(\omega')^2}{4} - \frac{\omega' \, b'}{b} = - (\phi')^2 \,,
\ee
and this is to be solved for $\omega$. In these variables the dilaton field equation \pref{Bdilatoneom2} similarly reads
\be
 \left( b\, e^\omega \, \phi' \right)' = \kappa^2 b \, e^\omega \left( \frac{2 g_\ssR^2}{\kappa^4} - \frac{ Q^2 }{ 2 } \, e^{-2\omega}  \right)   \,,
\ee
and flux quantization \pref{fluxqn} (for $N=1$ and $g_\ssA = g_\ssR$) can be written as
\be \label{eq:SIFQ}
 \frac{1}{g_\ssR} =  Q \int \d \sigma \, b \, e^{-\omega} - \frac{1}{2 \pi} \sum_\varv \zeta_\varv \,.
\ee

Finally, we rewrite the boundary conditions in the new variables, to get
\be
 \left[  b \, \phi' \right]_{\sigma_\varv} = \frac{\kappa^2 }{2 \pi} \Bigl[ T_\varv^\prime + Q \, \zeta_\varv^\prime \Bigr]_{\sigma_\varv} \,,
\ee
and
\be
 \left[ 1-  \left( b' - \frac12 \, b \, \phi' \right) \right]_{\sigma_\varv} =  \frac{\kappa^2 }{2 \pi} \Bigl[  T_\varv  \Bigr]_{\sigma_\varv} \,,
\ee
where $\sigma_\varv$ are the brane positions and the signs are such that the derivatives of $b$ and $\phi$ are in the direction away from the branes. In general the right-hand side of these boundary conditions generically diverge as $\sigma \to \sigma_\varv$. As shown explicitly in the examples of Appendix \ref{app:examples} this divergence can (and must) be renormalized into the parameters describing the brane-bulk couplings.

\subsubsection*{Linearized field equations}

The perturbations we consider to the tension and localized flux need not respect scale invariance, by depending nontrivially on the dilaton, and they can differ at each brane
\be \label{eq:deltaW}
 \zeta_\varv = \zeta_0 + \delta \zeta_\varv(\phi) \qquad \text{and} \qquad  T_\varv = T_0 + \delta T_\varv(\phi) \,.
\ee
The supersymmetric solutions are relatively simple because gradients in the warp factor and dilaton are absent: $\bar \omega = 0$ and $\bar \phi' = 0$. This need no longer be true given any asymmetry in the brane perturbations, and so to linearized order these bulks fields instead satisfy
\be
 \omega(\sigma) = \delta \omega(\sigma) \qquad \text{and} \qquad \phi'(\sigma) = \delta \phi'(\sigma) \,,
\ee
where primes again denote differentiation with respect to $\sigma$. These changes feed into the Einstein equations that govern the bulk metric function and the flux quantization condition that governs the size of $Q$, so
\be
 b(\sigma) = \bar b(\sigma) + \delta b(\sigma) \qquad \text{and} \qquad Q = \bar Q + \delta Q \,.
\ee

To solve for the field perturbations, we now linearize the full field equations around the supersymmetric rugby ball case. This gives the following Einstein equation for the metric function
\be
 \delta b''  +  \bar  b' \, \delta \omega' - \frac{1}{2} \left( \bar b \, \delta \phi' \right)' = - \frac{ \bar b }{\bar L^2} \left[ \frac{ \delta b }{ \bar b } +    \frac{3 }{2 } (\delta q -   \delta \omega) + \frac{1}{2} \kappa^2 \bar L^2 J e^{-\phi} \right] \,,
\ee
where $\delta q = \delta Q / \bar Q$. We also have the linearized dilaton field equation
\be \label{eq:lindilaton}
 \left( \bar b \, \delta\phi' \right)' =   \frac{ \bar b }{\bar L^2 }\left( \delta \omega - \delta q \right) \,.
\ee
Inserting this into the Einstein equation gives
\be \label{eq:linb}
 \delta b''  + \bar b' \, \delta \omega' = - \frac{\bar b}{\bar L^2} \left[ \frac{\delta b}{\bar b} + 2 ( \delta q - \delta \omega )  + \frac{1}{2} \kappa^2 \bar L^2 J e^{-\phi} \right] \,.
\ee
The linearized field equation for the warp factor simplifies a great deal
\be \label{eq:linwarp}
 \bar b  \, \delta \omega'' - \bar b' \, \delta \omega' = 0 \,.
\ee

The linearized boundary conditions reduce to
\be \label{eq:phibc}
 \left[ \bar b \, \delta \phi' \right]_{\bar \sigma_\varv} = \frac{\kappa^2 }{2 \pi} \Bigl[ \delta  T^\prime_\varv + \bar Q \, \delta \zeta_\varv^\prime  \Bigr]_{\bar \sigma_\varv} \,,
\ee
and
\be
\left[ \delta b'  + \frac12 \, \bar b \, \delta\phi' \right]_{\bar \sigma_\varv} = - \frac{\kappa^2 }{2 \pi } \Bigl[ \delta  T_\varv \Bigr]_{\bar \sigma_\varv} \,,
\ee
where $\bar \sigma_\varv = \{0,\pi \bar L \}$ are the unperturbed values of the brane positions, and again the sign assumes derivatives are directed away from the branes. Finally, combining the two boundary conditions gives an expression for the near-source derivative of the bulk metric function
\be \label{eq:bigBC}
 \left[   \delta b' \right]_{\bar \sigma_\varv} = - \frac{\kappa^2 }{2 \pi} \left[ \delta  T_\varv - \frac{1}{2} \delta  T^\prime_\varv - \frac{1}{2} \bar Q \, \delta \zeta_\varv^\prime  \right]_{\bar \sigma_\varv} \,.
\ee
In many of the above results we use the useful property of the unperturbed solution that $\kappa^2 \bar Q^2 \bar L^2 = 1$.

\subsubsection*{Linearized solutions}

The linearized field equations can all be solved analytically. Inserting the background solution into \pref{eq:linwarp} and integrating gives the following general solution for the warp factor
\be
 \delta \omega = \omega_0 + \omega_1 \cos z \,,
\ee
with $z := \sigma / \bar L$ and $\omega_0$ and $\omega_1$ both integration constants.

Absorbing the constant $\omega_0$ into a rescaling of the 4D coordinates and using the result in the linearized dilaton equation \pref{eq:lindilaton} then gives
\be
 \partial_z \left( \sin z \, \partial_z \delta \phi  \right) = \Bigl( \omega_1 \cos z - \delta q \Bigr) \sin z \,,
\ee
whose integral yields
\be \label{eq:phisoln}
 \sin z \, \partial_z  \delta \phi = -\frac{\omega_1 }{2} \, \cos^2 z + \delta q \cos z + \phi_1 \,,
\ee
where $\phi_1$ is another integration constant. Integrating again gives the full solution for the dilaton,
\be
 \phi = \bar \phi + \delta \phi = \varphi  - \frac{\omega_1}{2} \,  \cos z  + \left( \phi_1 - \frac{\omega_1}{2} \right) \log[\tan(z/2)] + \delta q \log(\sin z)  \,.
\ee

The constant part of the dilaton profile, $\varphi$, need not be perturbatively small so in $\phi$-dependent expressions that are already perturbatively small, like $J e^{-\phi}$, we can make the replacement $\phi \to \varphi$. This allows us to rewrite the Einstein equation \pref{eq:linb} as
\be
 \partial_z^2 \, \delta b   = -  \delta b  - 2 \bar \alpha \bar L \Bigl[ \delta q + \delta j(\varphi) \Bigr] \sin z + 3 \bar \alpha \bar L  \omega_1 \cos z \sin z \,,
\ee
where
\be
 \delta j(\varphi) := \frac{1}{4} \kappa^2 \bar L^2 J e^{-\varphi} \,.
\ee
The general solution to this field equation is given by
\be \label{eq:bsoln}
 \delta b = \bar \alpha \bar L \Bigl[ b_0 \cos z + b_1 \sin z  + (\delta q + \delta j)  z \cos z -  \omega_1 \sin z \cos z \Bigr] \,,
\ee
where $b_0$ and $b_1$ are integration constants. We are free to shift the radial coordinate to ensure $\delta b(0) = 0$ and thereby set $b_0 = 0$.

\subsubsection*{Changes in geometry}

The points $\sigma_\varv$ where the metric function vanishes define the brane positions. These are also perturbed relative to the background value $\sigma_\varv = \bar \sigma_\varv + \delta \sigma_\varv$ and we can solve for these perturbations by linearizing $b(\sigma_\varv) = 0$, which gives
\be
 0 = \delta b (\bar \sigma_\varv) + \delta \sigma_\varv \, [ \partial_\sigma \bar b ]_{\bar \sigma_\varv }  = \delta b (\bar \sigma_\varv) + \bar \alpha \delta \sigma_\varv \,.
\ee
This shows that the choice $b_0 = 0$ ensures that that one of the branes is always located at the origin $b(\sigma_0 ) = 0$ at linear order. At the other pole, near $\bar \sigma_\pi = \pi \bar L$, we instead find the following shift
\be
 \frac{ \delta \sigma_\pi }{\pi  \bar L} =  \delta q \,,
\ee
which shows how the backreaction can change the proper distance between the branes.

The change in scale invariant $k$-volume, defined as
\be
 \widehat  \Omega_{k} := 2 \pi \int \d \rho\, B W^k  e^{\phi}= 2 \pi \int \d \sigma \, b \, e^{k \omega/4} \,,
\ee
is given to linear order by the following expression
\be
 \delta \widehat \Omega_k = 2 \pi \int\limits_0^{\pi \bar L} \d \sigma \delta b +  \frac{ 2 \pi k}{4} \int\limits_0^{\pi \bar L} \d \sigma \, \bar b \, \delta \omega \,.
\ee
Evaluating the integral using the explicit solutions derived above gives
\be \label{eq:volumechange}
 \frac{ \delta \widehat  \Omega_k }{4 \pi \bar \alpha \bar L^2 } = \frac{1}{2} \int\limits_0^{\pi } \d z \Bigl[ b_1 \sin z  +  ( \delta q  + \delta j ) z \cos z  \Bigr] = b_1  -  \delta q  - \delta j \,,
\ee
where the integral over $\delta \omega$ vanishes because it is odd on the interval of integration. We learn the perturbation to the volume is independent of warping to linear order and is determined by the integration constants.

\subsubsection*{Boundary conditions and integration constants}

We next determine these integration constants in terms of the assumed brane perturbations, $\delta T_\varv$ and $\delta \zeta_\varv$, using the near-brane boundary conditions. We first evaluate the combined boundary condition \pref{eq:bigBC} using \pref{eq:bsoln} at both branes to get the following relation between integration constants and brane parameters
\be \label{eq:big}
    b_1 + \delta q + \delta j - \omega_1 e^{i  \varv} = - \frac{\kappa^2 }{2 \pi \bar \alpha} \left[ \delta  T_\varv - \frac{1}{2} \delta T^\prime_\varv - \frac{1}{2} \bar Q \, \delta \zeta_\varv^\prime  \right]_{\bar \sigma_\varv} \,,
\ee
where we use $\varv = \{ 0,\pi \}$ as an index to represent the branes located near $z = 0$ and $z = \pi$, and the explicit sign $e^{i  \varv}$ appears because the boundary conditions assume a radial coordinate that increases away from the brane (so the radial derivative in the boundary condition is $- \d/\d\sigma$ near $\sigma = \pi\bar L$ ). The dilaton boundary condition \pref{eq:phibc} similarly evaluates using the dilaton solution in \pref{eq:phisoln}, to give
\be \label{eq:phibceval}
 \delta q + \left( \phi_1  - \frac{\omega_1}{2} \right) e^{i \varv} = \frac{\kappa^2}{2 \pi \bar \alpha} \Bigl[ \delta  T_\varv^\prime + \bar Q \,\delta \zeta_\varv^\prime \Bigr]_{\bar \sigma_\varv} \,.
\ee

The integration constant controlling the gradient in $W$ is fixed by the difference between the \pref{eq:big} at $\varv =0$ and $\varv = \pi$ in terms of brane differences
\be
 \omega_1 = \frac{\kappa^2 }{2 \pi \bar \alpha} \left( \delta  T_{\rm dif} - \frac{1}{2} \delta  T^\prime_{\rm dif} - \frac{1}{2} \bar Q \, \delta \zeta_{\rm dif}^\prime \right)  \,,
\ee
where $\delta  T_{\rm dif} = \frac{1}{2}( \delta  T_{\varv=0} - \delta  T_{\varv=\pi} )$ and so on. Using this in the difference between the two versions of \pref{eq:phibceval} similarly determines the gradient of $\phi$ by fixing
\be
 \phi_1  = \frac{\kappa^2}{2 \pi \bar \alpha} \left( \frac{1}{2} \delta  T_{\rm dif}+  \frac{3}{4} \delta  T_{\rm dif}^\prime + \frac{3}{4} \bar Q \, \delta \zeta_{\rm dif}^\prime \right) \,.
\ee

The remaining integration constants are found by summing rather than subtracting boundary conditions, and the two versions of \pref{eq:phibceval} sum to give $\delta q$ in terms of brane averages
\be \label{eq:deltaqsolved}
 \delta q = \frac{\kappa^2}{2 \pi \bar \alpha } \left( \delta  T_{\rm avg}^\prime + \bar Q \,\delta \zeta_{\rm avg}^\prime \right) \,,
\ee
where $\delta  T_{\rm avg} = \frac{1}{2} ( \delta  T_{\varv=0} + \delta  T_{\varv=\pi} )$ and so on. This expression can be used in conjunction with \pref{eq:big} to solve for the last integration constant, $b_1$, and we find
\be \label{eq:firstb1}
 b_1 + \delta j = - \frac{\kappa^2 }{2 \pi \bar \alpha } \left( \delta  T_{\rm avg} + \frac{1}{2} \delta  T^\prime_{\rm avg} + \frac{1}{2} \bar Q\, \delta \zeta_{\rm avg}^\prime  \right) \,.
\ee
These four conditions completely fix the integration constants, $\phi_1$, $\omega_1$, $b_1$ and $\delta q$ in terms of the brane parameters, the stabilizing current $J$ and $\varphi$ (which to this point remains arbitrary).

A final relation comes from the linearized flux-quantization condition,
\be
 0 = 4 \pi \alpha \bar L^2 \, \delta Q + \bar Q \delta \widehat \Omega - \sum_\varv \delta \zeta_\varv  \,,
\ee
which we use to determine $J$ in terms of brane properties and $\varphi$. To this end we use the linearized volume change in \pref{eq:volumechange} to fix $b_1$
\be \label{eq:moreb1}
 b_1 =  \frac{\kappa^2}{ 2 \pi \bar \alpha }  \left( \bar Q\, \delta \zeta_{\rm avg} \right)\,,
\ee
where we have used $\kappa^2 \bar L^2 \bar Q^2 =1$ to write the linearized flux quantization condition in this suggestive manner. Combining \pref{eq:firstb1} with \pref{eq:moreb1} gives a solution for the stabilizing current
\be \label{eq:branerule}
 \delta j  = \frac{1}{4} \kappa^2 \bar L^2 J e^{-\varphi} = - \frac{\kappa^2} {2 \pi \bar \alpha} \left( \delta  T_{\rm avg}  + \bar Q \delta \zeta_{\rm avg}  + \frac{1}{2} \delta  T^\prime_{\rm avg} + \frac{1}{2} \bar Q \delta \zeta^\prime_{\rm avg} \right) \,.
\ee
Setting $J = 0$ in this gives the stabilized value, $\varphi = \varphi_\star$, of the zero mode entirely in terms of the brane parameters.

\subsubsection*{The effective potential}

We now construct the effective potential of the 4D theory using the ancillary current $J$, and verify that it is minimized by the condition \pref{eq:branerule}.
The addition of $\Delta S_\ssJ$ to the bulk action gives rise to a corresponding term in the effective theory. To identify how the current contributes to the effective theory we note that it can be treated like a novel contribution to the 6D potential
\be
 \Delta V_\ssB = J  \,.
\ee
Using this in \pref{eq:UofVB} shows that the presence of a stabilizing current in the 6D theory can be captured by shifting the overall potential in the 4D theory as follows
\be
 \Delta U = \frac{1}{2} e^{2 (\varphi - \varphi_\star)} \la J \ra = \frac{1}{2} e^{2 (\varphi - \varphi_\star)} \int \d^2 y \sqrt{\hat g_2} \, J e^{-\phi} \,.
\ee
To the linear order of interest this gives
\be
 \Delta U = 2 \pi \bar \alpha \bar L^2 \, e^{2( \varphi - \varphi_\star)}  \, J e^{-\varphi } \,,
\ee
where the dependence on $\varphi$ only appears in in the exponential factors. The presence of this additional term in the 4D theory modifies the field equation for the zero mode
\be
 \frac{ \partial U}{\partial \varphi} + 2 \pi \bar \alpha \bar L^2 \, e^{2(\varphi- \varphi_\star)}  \, J e^{- \varphi} = 0 \,.
\ee

Since $J$ is a known function of $\varphi$ this can be read as a differential equation for the potential which is solved by
\be
 U(\varphi) =  - 2 \pi \bar \alpha \bar L^2 \, e^{-2 \varphi_\star} \int \d \tilde \varphi \, J  e^{\tilde \varphi} \,.
\ee
This can be combined to with \pref{eq:branerule} to determine the potential in terms of brane perturbations
\be
 U(\varphi) = 4 e^{-2 \varphi_\star} \int \d \tilde \varphi e^{2 \tilde \varphi} \left( \delta  T_{\rm avg}  + \bar Q \delta \zeta_{\rm avg}  + \frac{1}{2} \delta  T^\prime_{\rm avg} + \frac{1}{2} \bar Q \delta \zeta^\prime_{\rm avg} \right) \,.
\ee
It is possible to directly integrate this expression to find the linearized potential
\be \label{eq:finalpotential}
 U(\varphi) = 2 \, e^{2(\varphi - \varphi_\star )} \left[   \delta  T_{\rm avg} + \left( \frac{2 g_\ssR}{\kappa^2} \right) \delta \zeta_ {\rm avg} \right] \,,
\ee
where we have used $\bar Q = 2 g_\ssR / \kappa^2.$ There is no background contribution to the potential because it vanishes identically in the supersymmetric case around which we are perturbing.

This potential agrees with the linearization of the potential found by dimensional reduction in \S\ref{phidependence} of the main text, and correctly predicts that the energy is perturbed by $\sum_\varv \delta T_\varv$ at linear order when the brane tension is perturbed in a way that vanishes when the brane perturbations are scale invariant and supersymmetric. Furthermore, minimizing the potential for general brane perturbations gives the same condition on the zero mode as \pref{eq:branerule} gives when $J=0$, and this confirms that the effective potential reproduces the stabilization of the zero mode that was derived in the 6D theory.

\section{Examples of stabilization}
\label{app:examples}

We now investigate simple examples of zero mode stabilization by choosing explicit forms for the $\phi$-dependence of the brane perturbations. In all cases, we imagine the $\phi$-dependence of the brane appears predominantly in the flux perturbation, since we expect this choice to help suppress vacuum energies, as in \pref{eq:Udecouple}. In many cases we find that exponentially large extra dimensions and suppressed curvatures can be obtained if there is a hierarchy between the size of the brane perturbations.

Along the way, we also illustrate how classical renormalization of brane parameters can be used to absorb divergences that arise when the brane is treated as an idealized, infinitely thin source. The procedure renders finite physical observables like the value of the zero mode, and the potential at its minimum.

\subsubsection*{Flux with linear $\phi$-dependence}

We now investigate simple example in which the the branes are perturbed identically, with the following properties
\be
 \delta  T_\varv = \tau \qquad {\rm and} \qquad \delta \zeta_\varv = \lambda \, \phi \,.
\ee
This choice of identical brane perturbations immediately gives
\be
\phi_1 = \omega_1 = 0 \,.
\ee
Note that this greatly simplifies the solution for the dilaton
\be
 \phi = \varphi  + \delta q \log \left( \sin z \right) \,.
\ee
The remaining unknown integration constant can be calculated from \pref{eq:deltaqsolved} and it gives $\delta q = \lambda g_\ssR / \pi \bar \alpha.$ Inserting this into \pref{eq:branerule} and setting $J=0$ gives an equation to be solved for the value of the zero mode
\be
 0 = 2 \lambda g_\ssR \varphi_\star +  \left( \frac{ 2 \lambda^2 g_\ssR^2}{ \pi \bar \alpha} \right) \log(\epsilon/\bar L)   + \kappa^2 \tau +  \lambda g_\ssR   \,.
\ee
Note that we have regularized the $\lim_{\sigma \to 0} \log[\sin(\sigma/\bar L)]$ divergence with the finite expression $\log(\epsilon/\bar L).$ However, the divergence as $\epsilon \to 0$ must be absorbed into the brane couplings such that physical quantities are finite.

In general, divergences associated with brane terms that are linear in a bulk scalar can be absorbed by renormalizing the $\phi$-independent part of the brane tension \cite{6DHiggsStab}. This case is no different, and the observables of the theory can be made finite if we renormalize the tension as follows
\be
 \kappa^2 \tau(\bar r) = \kappa^2 \tau - \left( \frac{2 \lambda^2 g_\ssR^2 }{ \pi  \bar \alpha} \right) \log(\epsilon / \bar r) \,.
\ee
In particular, this renormalization gives a finite expression for the value of the zero mode
\be
 \varphi_\star  = -  \left[ \frac{ \kappa^2 \tau(\bar r)}{2 \lambda g_\ssR} \right] - \frac{1}{2} - \left( \frac{\lambda g_\ssR}{ \pi \bar \alpha} \right) \log( \bar r / \bar L ) \,.
\ee
For convenience, we can choose the renormalization scale $\bar r = \bar L$ to eliminate the logarithmic term and this gives
\be \label{eq:linearvarphi}
  \varphi_\star = -  \left[ \frac{ \kappa^2 \tau(\bar L)}{2 \lambda g_\ssR} \right] - \frac{1}{2} \,.
\ee
Note that the limit $\lambda \to 0$ sends to zero mode to the expected runaway value $\varphi_\star \to - \infty.$ Also note that the value of the zero mode comes to us as the ratio of two small, dimensionless numbers $\kappa^2 \tau(\bar L)$ and  $\lambda g_\ssR$ but can itself be made large if $\lambda g_\ssR \ll \kappa^2 \tau(\bar L).$ Because the proper volume of extra dimensions is controlled by $\ell^2 = r_\ssB^2 e^{-\varphi_\star}$, a large negative value of the zero mode gives large extra dimensions. Furthermore, this choice does not invalidate the assumed perturbativity of $\kappa^2 \bar Q \delta \zeta \approx  \lambda g_\ssR \varphi \approx \kappa^2 \tau(\bar L) \ll 1 $ near the minimum of the potential, and so the approximate, linearized potential is valid in this region.

We can therefore calculate this potential, and using \pref{eq:finalpotential} gives
\be
 U(\varphi) =\frac{ 2 \, e^{2(\varphi - \varphi_\star )} }{\kappa^2} \left[   \kappa^2 \tau   + 2 \lambda g_\ssR  \varphi  +  \left( \frac{2 \lambda^2 g_\ssR^2}{\pi \bar \alpha} \right) \log(\epsilon/\bar L) \right] \,.
\ee
Again, the logarithmic divergence of the scalar field has been regularized with a finite regulator $\epsilon.$ Conveniently, though not surprisingly, this divergence can be renormalized into the tension to yield a finite potential
\be
U = \frac{ 2 \, e^{2(\varphi - \varphi_\star )} }{\kappa^2} \left[   \kappa^2 \tau  + 2 \lambda g_\ssR  \phi(\bar \sigma_0) \right] = \frac{ 2 \, e^{2(\varphi - \varphi_\star )} }{\kappa^2} \left[   \kappa^2 \tau(\bar L)   + 2 \lambda g_\ssR  \varphi   \right] \,,
\ee
where we have chosen the renormalization scale $\bar r = \bar L$ so that the finite logarithms are all implicit. Minimizing this potential reproduces the solution for the zero mode in \pref{eq:linearvarphi} as it should. Finally, the value of the potential at the minimum is
\be
 U_\star = - 2\tau \left( \frac{\lambda g_\ssR}{\kappa^2 \tau} \right) \,.
\ee
In the parameter range $\kappa^2 \tau(\bar L)/ \lambda g_\ssR  \gg 1$ that gives large dimensions, the vacuum energy is suppressed relative to the naive expectation $2 \tau$.

\subsubsection*{Flux quadratic in $\phi$}

We now consider the case in which the perturbation to the localized flux is quadratic in $\phi$
\be
 \delta  T_\varv = \tau \qquad \text{and} \qquad \delta \zeta_\varv = m \, \phi^2 \,.
\ee
We again make the simplifying assumption of identical branes and this gives $\phi_1 = \omega_1 = 0$ so that $\phi = \delta q \log(\sin z) + \varphi .$ Inserting this into \pref{eq:deltaqsolved} allows us to rewrite it as follows
\be
 \delta q = \frac{2 m g_\ssR}{\pi \bar \alpha} \left[ \delta q \log( \epsilon/\bar L) + \varphi \right] \,,
\ee
where the logarithmic divergence of the dilaton is $\epsilon$-regularized in the same way as before. This equation can be used to solve for $\delta q$ in terms of the zero mode
\be
 \pi \bar \alpha \delta q = \frac{ 2 m g_\ssR  \varphi }{1 - \frac{ 2 m g_\ssR }{ \pi \bar \alpha} \log(\epsilon/\bar L) } \,.
\ee
When the brane has a quadratic coupling to a bulk scalar field, the associated divergences can be absorbed into the renormalization of this this coupling's coefficient \cite{6DHiggsStab, ClassRenorm}. In the present case this amounts to renormalizing $m$ as follows
\be
 m(\bar r) = \frac{m}{1 - \frac{2 m g_\ssR  }{\pi \bar \alpha} \log(\bar \epsilon / \bar r)} \,.
\ee
This gives a finite value for the $\delta q$ as a function of the zero mode $\pi \bar \alpha \delta q =2  g m(\bar L) \varphi$ because it absorbs the divergences associated with evaluating the dilaton profile at the brane positions $m \phi(\bar \sigma_0 ) = m(\bar L) \varphi.$

The condition on the zero mode in \pref{eq:branerule} is also finite after renormalization and approriately linearizing $\phi^2.$ It reads
\be
 0 = \kappa^2 \tau +  2 g_\ssR m(\bar L) \varphi_\star + 2 g_\ssR m(\bar L) \varphi_\star^2 \,.
\ee
This can alternatively be derived by minimizing the linearized and renormalized potential
\be
U(\varphi) = \frac{ 2 e^{2(\varphi - \varphi_\star) } }{\kappa^2} \left[ \kappa^2 \tau +2  g_\ssR m \phi^2(\bar \sigma_\varv) \right] = \frac{ 2 e^{2(\varphi - \varphi_\star) } }{\kappa^2} \left[ \kappa^2 \tau + 2 g_\ssR m(\bar L)  \, \varphi^2 \right] \,,
\ee
where we have used $m \phi^2(\bar \sigma_\varv) = m(\bar L) \varphi \phi = m(\bar L) \varphi^2$ to linear order. The solution for the zero mode reads
\be
 \varphi_\star = \frac{1}{2} \left( \pm \sqrt{1 +  2 t } - 1 \right)
\ee
where $t = -\kappa^2 \tau / g m(\bar L).$ There are real roots when $ t \ge -1/2 $ and they are both negative unless $t > 0$ at which point one of them switches sign. The concavity of the potential at the extrema is proportional to $g m(\bar L) \left( \varphi_\star  + \frac{1}{2} \right)  = \pm g m(\bar L)  \sqrt{1 + 2 t}$. So if there are extrema, one of them is always a minimum and the other is always a maximum. The minimum can occur at the more negative root if $g m(\bar L)$ is also negative.

If we assume $t \gg 1$ then the stabilized value of the zero mode is dominated by the root of the large ratio $t$ as follows
\be
 \varphi_\star = \pm \sqrt{ \left| \frac{ \kappa^2 \tau}{ 2 g_\ssR m(\bar L)} \right| } \,,
\ee
and the negative solution can be a minimum if $g_\ssR m(\bar L) < 0.$ This would also require $\kappa^2 \tau >0$ if $t > 0$ is to be satisfied. If the stabilized value of $\varphi_\star$ is chosen to be large and negative, then the extra dimensions have exponentially large radius as suggested by the leading order result $\ell^2 = r_\ssB^2 e^{- \varphi_\star}.$

Finally, the value of the potential at this minimum can be written as follows
\be
 U_\star = - \left( \frac{4 g_\ssR }{\kappa^2}   \right) m(\bar L) \varphi_\star = - 2 \tau  \sqrt{\frac{ 2 |g_\ssR m(\bar L)| }{ \kappa^2 \tau } }\,.
\ee
Similar to the linear case, this vacuum energy is suppressed relative to the naive expectation $2 \tau$, though the suppression here is weaker because it is sensitive to the root of the hierarchy in the brane perturbations.

\end{document}